\documentclass[11pt,a4paper]{article}
\pdfoutput=1
\usepackage{setspace}
\usepackage{jheppub}
\usepackage{rotating}
\usepackage{array}
\usepackage{amsmath}
\usepackage[normalem]{ulem}
\usepackage{slashed}
\usepackage{booktabs}
\usepackage[pdftex,table,dvipsnames]{xcolor}
\usepackage{units}
\usepackage{multirow}
\usepackage{url}
\usepackage{xspace}
\usepackage{color}
\usepackage{float}
\usepackage{lipsum}
\usepackage[utf8]{inputenc}
\DeclareUnicodeCharacter{2032}{\ensuremath{^{\prime}}}
\newcommand{\rund}[1]{\left( #1 \right)}
\newcommand{\eck}[1]{\left\lbrack #1 \right\rbrack}

\newcommand{\mll}{m_{\ell \ell}}
\newcommand{\gqv}{g_q^V}
\newcommand{\glv}{g_{\ell}^V}
\newcommand{\gqa}{g_q^A}
\newcommand{\gla}{g_{\ell}^A}
\newcommand{\gfv}{g_f^V}
\newcommand{\gfa}{g_f^A}

\newcommand{\gxv}{g_{\chi}^V}

\newcommand{\ps}{\hat{s}} 
 
\newcommand{\Tf}[2]{\mathcal{T}^{\, #1}_{\, #2} }

\renewcommand{\d}{\mathrm{d}}
\newcommand{\Zp}{Z^{\prime}}
\newcommand{\mz}{m_{Z^{\prime}}}
\newcommand{\mx}{m_{\chi}}
\newcommand{\gz}{\Gamma_{Z^{\prime}}}
 
\newcommand{\delphes}{\texttt{DELPHES}} 

\newcommand{\citere}[1]{Ref.~\cite{#1}}
\newcommand{\citeres}[1]{Refs.~\cite{#1}}
\newcommand{\eqn}[1]{eq.~(\ref{#1})}

\newcommand{\tab}[1]{table~\ref{#1}}
\newcommand{\fig}[1]{figure~\ref{#1}}

\preprint{ TTK-19-53, MPP-2019-253, P3H-19-059}

\title {Interference effects in dilepton resonance searches \\ for $Z'$ bosons and dark matter mediators}

\author[a]{Felix Kahlhoefer,}
\author[a]{Alexander M{\"u}ck,}
\author[a,b]{Stefan Schulte,}
\author[a]{and Patrick Tunney}

\affiliation[a]{Institute for Theoretical Particle Physics and Cosmology (TTK), RWTH Aachen University, \\ D-52056 Aachen, Germany}
\affiliation[b]{Max Planck Institute for Physics, F{\"o}hringer Ring 6, D-80805 M{\"u}nchen, Germany}

\emailAdd{kahlhoefer@physik.rwth-aachen.de}
\emailAdd{mueck@physik.rwth-aachen.de}
\emailAdd{sschulte@mpp.mpg.de}
\emailAdd{tunney@physik.rwth-aachen.de}

\abstract{New $\Zp$ gauge bosons arise in many extensions of the Standard Model and predict resonances in the dilepton invariant mass spectrum. Searches for such resonances therefore provide important constraints on many models of new physics, but the resulting bounds are often calculated without interference effects. In this work we show that  the effect of interference is significant and cannot be neglected whenever the $\Zp$ width is large (for example because of an invisible contribution). To illustrate this point, we implement and validate the most recent $139\,\mathrm{fb^{-1}}$ dilepton search from ATLAS and obtain exclusion limits on general $\Zp$ models as well as on simplified dark matter models with spin-1 mediators. We find  that interference can substantially strengthen the bound on the $\Zp$ couplings and push exclusion limits for dark matter simplified models to higher values of the $\Zp$ mass. Together with this study we release the open-source code \texttt{ZPEED}, which provides fast likelihoods and exclusion bounds for general $\Zp$ models.}


\begin{document}

\maketitle

\flushbottom

\section{Introduction}

The dijet and dilepton final states are amongst the simplest channels currently considered by the LHC collaborations. While dijet resonance searches~\cite{Aad:2019hjw,Sirunyan:2019vgj} have the advantage that any new particle produced from $q\bar{q}$ annihilation necessarily can decay back into a pair of quarks, searching for an excess can be difficult due to the large QCD background. Instead, dilepton searches look for a similar bump-like feature above a much smaller electroweak background and achieve great sensitivity to any new particle that couples to Standard Model (SM) leptons~\cite{Aaboud:2017buh, Sirunyan:2018exx, Aad:2019fac, CMS:2019tbu}.

In particular, these searches for exotic resonances offer us a powerful way to probe theories with a new spin-one mediator $\Zp$. Such $\Zp$ bosons generically appear in many extensions of the SM~\cite{An:2012va,An:2012ue,Accomando:2013sfa,Belyaev:2013xfa,Accomando:2010fz,Basso:2010pe,Accomando:2019ahs,Altarelli:1989ff,Langacker:2008yv,Kim:2014afa,Basso:2012ux,Accomando:2016mvz,Accomando:2017fmb,Alves:2016cqf,Gulov:2018zij} and are an essential part of Grand Unified Theories~\cite{Langacker:1980js,Hewett:1988xc}. On a more phenomenological level, they have also received substantial attention as the mediator of spin-one simplified models of Dark Matter (DM)~\cite{Malik:2014ggr,Abdallah:2015ter,Bauer:2016gys,Frandsen:2012rk,Fox:2012ru,Alves:2013tqa,Arcadi:2013qia,Buchmueller:2014yoa,Lebedev:2014bba,Harris:2014hga,Busoni:2014gta,Fairbairn:2014aqa,Altmannshofer:2014cla,Jacques:2015zha,Alves:2015pea,Chala:2015ama,DeSimone:2016fbz,Brennan:2016xjh,Jacques:2016dqz,Capdevilla:2017doz,Blanco:2019hah}. These simplified models have been advertised by the LHC DM working group~\cite{Boveia:2016mrp,Albert:2017onk} in order to explore the complementarity between different LHC analyses and across different DM experiments, which include direct and indirect detection, as well as observations of the DM relic density.

While originally these simplified models focused exclusively on the interactions between DM and quarks, it was soon pointed out that lepton couplings cannot be neglected. In models where the $\Zp$ couples differently to left- and right-handed quarks, the presence of lepton couplings is imposed both by considerations of gauge invariance and by the requirement that there are no gauge anomalies (assuming no exotic SU(2) fermions or additional Higgs doublets). But even in models with vector-like couplings to quarks, lepton couplings generally arise through loop-induced kinetic mixing. Searches for dilepton resonances therefore often place the strongest constraints on simplified DM models and in many cases exclude the most interesting regions of parameter space~\cite{Kahlhoefer:2015bea,Duerr:2016tmh,Duerr:2017uap,Ellis:2017tkh,Ellis:2018xal,Caron:2018yzp,ElHedri:2018cdm}.

In this work, we point out that existing bounds on simplified DM models from dilepton resonance searches are inaccurate, because they neglect the effect of interference between the $\Zp$ signal and the SM Drell-Yan background $pp \to Z^\ast / \gamma^\ast \to \ell^+ \ell^-$~\cite{Accomando:2019ahs,Accomando:2013sfa} (see also Ref.~\cite{Raj:2016aky}). It is commonly assumed that the impact of this interference is negligible, which is typically a good approximation for narrow resonances. However, in the context of simplified DM models, this assumption is not justified because decays of the $\Zp$ into DM particles can give a large additional contribution to the width of the $\Zp$, called the \emph{invisible width}. If the size of the DM coupling is larger than the SM couplings, the width of the $\Zp$ will significantly increase as the phase space for the invisible decay opens up.

We demonstrate that the effect of interference can be large, in particular if the signal is smaller than the background and spread out across several bins. In particular for small $\Zp$ masses ($\mz < 2 \, \mathrm{TeV}$) and large widths ($\gz / \mz > 3\,\%$), upper bounds on the couplings can improve by up to a factor of 1.5. The code used to obtain these results is publicly available and can be downloaded from \url{https://github.com/kahlhoefer/ZPEED}. In the interest of computational speed, the code makes essentially no use of Monte Carlo event generators, relying instead on analytical cross section calculations and exploiting that the effects of parton distribution functions (PDFs) and analysis cuts are essentially model-independent.

Our paper structure is then as follows. Section~\ref{sec:Interference} introduces our calculation of cross sections for dilepton processes and shows the effect of interference on signal shapes. In section~\ref{sec:Analysis} we describe our implementation of an ATLAS search for dilepton resonances with $139\,\mathrm{fb^{-1}}$ of data \cite{Aad:2019fac}, including the modeling of detector effects, the statistical method, and a validation via comparison to published bounds. In section~\ref{sec:DM} we then present our results on the importance of interference effects for various $\Zp$ models, with a special focus on a DM simplified model, with benchmark couplings recently proposed by the LHC DM working group \cite{Albert:2017onk}. Finally, our conclusions are presented in section~\ref{sec:Conclusion}.

\section{Interference effects for vector resonances}
\label{sec:Interference}

\begin{figure}[t]
\centering
\includegraphics[scale=1]{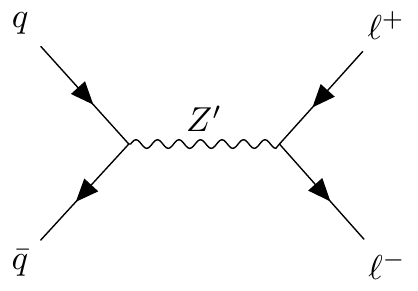}
\vspace{-2mm}
\caption{Feynman diagram for $s$-channel annihilation of a quark-antiquark pair into leptons mediated by a $Z'$ at leading order. Diagram created with \texttt{TikZ-Feynman}~\cite{Ellis:2016jkw}.} \label{fig:FeynmanDilepton}
\vspace{-2mm}
\end{figure}

In this section we describe the calculation of the cross section for $pp \to \Zp \to \ell^+ \ell^-$ (see figure \ref{fig:FeynmanDilepton}) at leading order, including the effect of interference with the SM background processes mediated by the $Z$ boson and the photon. This issue has previously been studied in the context of specific $\Zp$ models in~\citere{Accomando:2013sfa}. For this purpose we introduce a generic $\Zp$ model with the interaction Lagrangian
\begin{equation}
\label{eq:Lagrangian}
\mathcal{L}_\text{int} = - \sum_{f} \, Z^{\prime\mu} \, \bar{f} \, \eck{g_f^V \gamma_{\mu} + g_f^A \gamma_{\mu}\gamma^5} \, f  \; ,
\end{equation} 
where $\Zp$ is the spin-one mediator (with mass $\mz$), $f$ is a SM fermion and $g^{V/A}$ are vectorial/axial couplings. Since we wish to remain agnostic about the possible existence of additional contributions to the total width, we treat the decay width of our $\Zp$, denoted by $\gz$, as a free parameter in this section.
The cross section for the full hadronic process can be related to the partonic one for the hard process as
\begin{align}
\label{eq:GeneralXS}
\sigma(pp \longrightarrow \ell^+ \ell^-) & = \sum_q \int \d{x_1}\d{x_2} \;  f_q (x_1) f_{\bar{q}} (x_2) \; \hat{\sigma}(q\bar{q} \longrightarrow \ell^+ \ell^-) \; , 
\end{align}
where the sum is performed over all quark and anti-quark flavours. Here, the $x_i$ denote the momentum fractions of the individual partons and $f_q$ and $f_{\bar{q}}$ are the \texttt{MSTW} PDFs~\cite{Martin:2009iq}, which we evaluate setting the factorisation scale to $\mu = \mll$. It is straight-forward from this expression to calculate the differential cross section with respect to the dilepton invariant mass $\mathrm{d}\sigma / \mathrm{d}\mll$ (see appendix~\ref{app:xsec}). This cross section can be split into several parts:
\begin{equation}
\label{eq:SigmaHatSplit}
\frac{\mathrm{d}\sigma}{\mathrm{d}\mll} =  \underbrace{\frac{\mathrm{d}\sigma_{\gamma\gamma}}{\mathrm{d}\mll} + \frac{\mathrm{d}\sigma_{ZZ}}{\mathrm{d}\mll}+ 2\cdot \frac{\mathrm{d}\sigma_{\gamma Z}}{\mathrm{d}\mll}}_{\mathrm{d}\sigma_\text{background}/\mathrm{d}\mll} + \underbrace{\vphantom{+ \frac{\mathrm{d}\sigma_{\gamma Z}}{\mathrm{d}\mll}} \frac{\mathrm{d}\sigma_{\Zp\Zp}}{\mathrm{d}\mll}}_{\mathrm{d}\sigma_\text{signal}/\mathrm{d}\mll}+ \underbrace{ 2\cdot \frac{\mathrm{d}\sigma_{\Zp\gamma}}{\mathrm{d}\mll} +  2\cdot \frac{\mathrm{d}\sigma_{\Zp Z}}{\mathrm{d}\mll}}_{\mathrm{d}\sigma_\text{interference}/\mathrm{d}\mll} \; .
\end{equation}

In dilepton resonance searches the SM background is typically large (at least for $\mll \lesssim 2\,\mathrm{TeV}$) but known with a high level of precision. These searches are therefore potentially sensitive to exotic resonances even if in any given bin $\sigma_\text{signal} \ll \sigma_\text{background}$.\footnote{Here we define $\sigma = \int_a^b (\mathrm{d}\sigma / \mathrm{d} \mll) \mathrm{d} \mll$ for a bin given by $\mll \in [a, b]$. Typical bin sizes are comparable to the detector resolution, which is approximately 1--2\% in the electron channel and 5--10\% in the muon channel. Note that for the purpose of this section we neglect detector effects, which will be discussed in detail in section~\ref{sec:Detector}.} For many $\Zp$ models the width $\gz$ is small compared to the bin size. In this case the signal will only be observable if $\mathrm{d}\sigma_\text{signal}/\mathrm{d}\mll \gg \mathrm{d}\sigma_\text{background}/\mathrm{d}\mll$ for $\mll \approx \mz$. Since
\begin{equation}
 \left(\frac{\mathrm{d}\sigma_\text{interference}}{\mathrm{d}\mll}\right)^2 < 4 \frac{\mathrm{d}\sigma_\text{signal}}{\mathrm{d}\mll} \frac{\mathrm{d}\sigma_\text{background}}{\mathrm{d}\mll} \; ,
\end{equation}
it follows that $\mathrm{d}\sigma_\text{signal}/\mathrm{d}\mll \gg \mathrm{d}\sigma_\text{interference}/\mathrm{d}\mll$, so that interference effects are typically not important. If on the other hand $\gz$ is comparable to the bin size (for example because of an invisible decay mode), dilepton resonance searches are potentially sensitive to signals with $\mathrm{d}\sigma_\text{signal}/\mathrm{d}\mll \ll \mathrm{d}\sigma_\text{background}/\mathrm{d}\mll$ for all values of $\mll$. For such small signal cross sections, interference effects can potentially be very important.

\begin{figure}[t]
    \centering
        \includegraphics[width=0.49\textwidth]{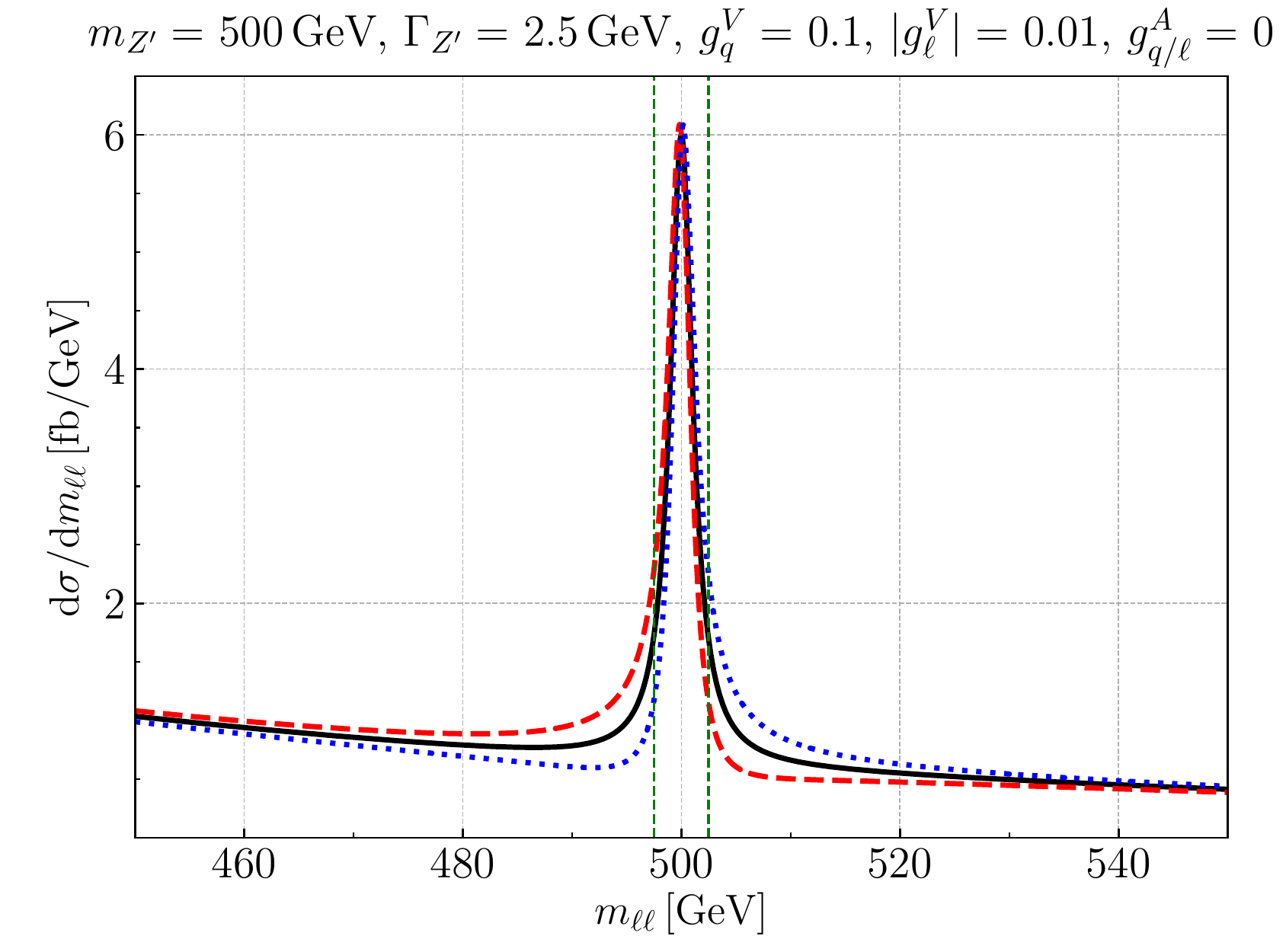}
        \includegraphics[width=0.49\textwidth]{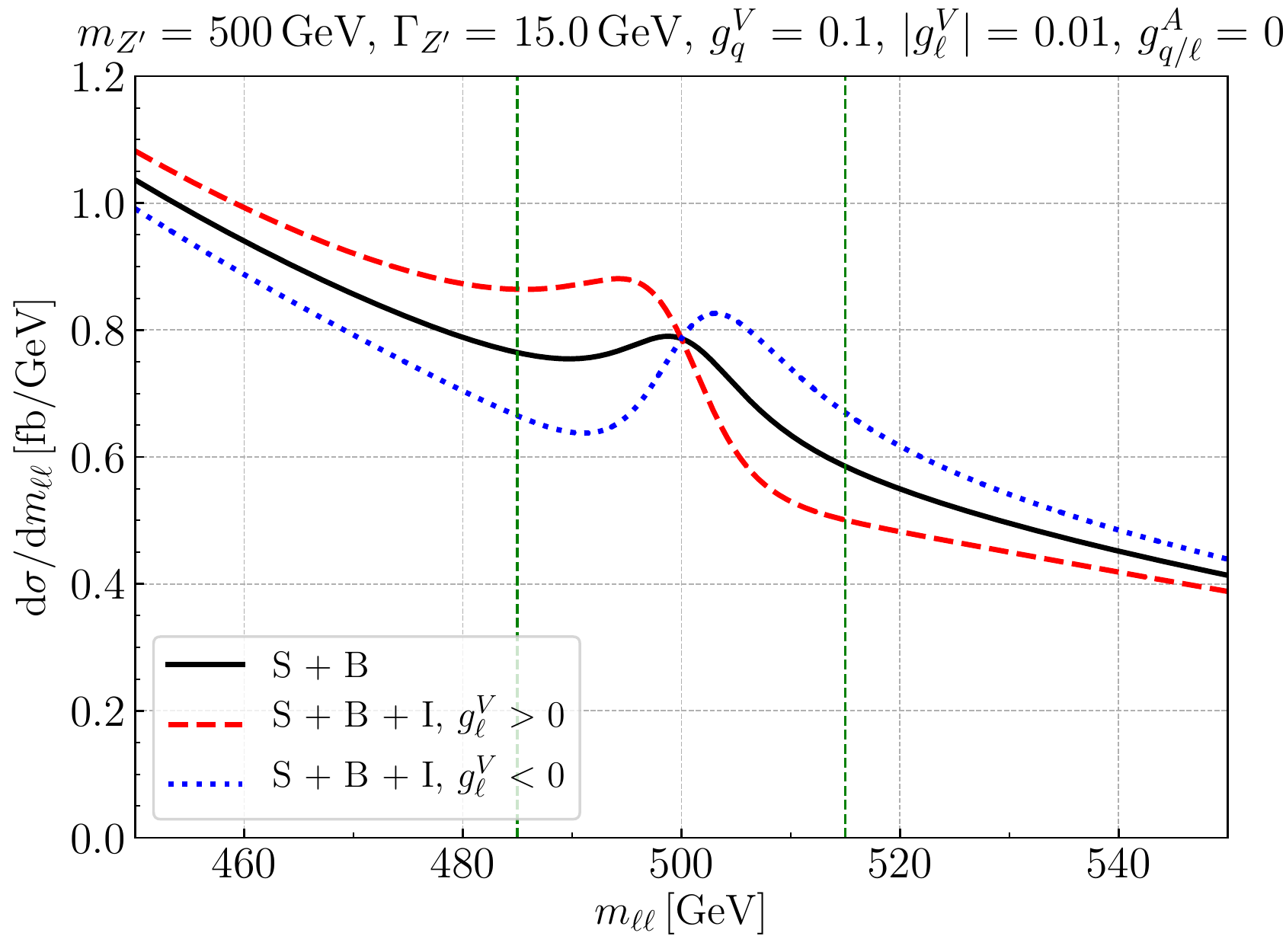}
        \vspace{-2mm}
    \caption{Comparison of the spectrum $\d{\sigma}/\d{\mll}$ as function of $\mll$ for different decay widths. The left panel displays the differential cross section for $\gz = 2.5\,$GeV while the right panel shows $\gz = 15\,$GeV. We show a naive addition of signal (S) and background (B) as a black line, and the full result with interference (I) included as blue dotted and red dashed lines for both signs of $\glv$. In the right panel, the signal shape in the region $\lbrack   \mz-\gz ,\, \mz+\gz \rbrack$ (indicated by the dashed green lines) is clearly affected by the inclusion of interference terms. }
    
        \vspace{-2mm}
\label{fig:InterferenceSignals}
\end{figure}

We illustrate the effect of interference in \fig{fig:InterferenceSignals} for a narrow $\Zp$ signal with width $2.5\,$GeV~(left panel) and a broad signal with width $15\,$GeV~(right panel), keeping the couplings and resonance mass fixed.\footnote{These widths approximately correspond to the minimal width from decays into SM states and the width with a light DM fermion included, respectively, for the model that we will consider in section~\ref{sec:DM}.} The differential cross section as a function of $\mll$ is shown for the naive sum of signal and background without interference and with interference included.

As expected, we find that interference becomes more important for larger widths, because of the suppression of the pure signal term compared to the interference term. In detail the effect of interference depends on the sign of the $\Zp$ couplings. For $\gqv \glv > 0$ interference is constructive for $\mll < \mz$ and destructive for $\mll > \mz$. Since the background is monotonically falling, this leads to an increase in the height of the peak and a shift of its location to smaller values of $\mll$. For the opposite case ($\gqv \glv < 0$) the height of the peak still increases, but the peak is now shifted to larger values of $\mll$. In the following, we will focus on the case that $\gqv \glv > 0$. Results for the opposite case are summarized in appendix~\ref{app:opposite}.

To conclude this section, we note that interference effects are more relevant for $\Zp$ mediators with vector couplings than for those with axial couplings. The reason is that axial mediators do not interfere with the photon, which gives the dominant contribution to interference for vector mediators. We will therefore restrict ourselves to vector mediators in the following.

\section{Analysis set-up}
\label{sec:Analysis}

In this section we describe how to translate the theoretical cross section from above into realistic predictions of the expected number of events in a given set of bins of the 
dilepton invariant mass $\mll$, including analysis cuts, detector efficiencies, energy resolution and higher-order effects. We then give a brief summary of the statistical method that we employ to test whether or not the resulting signal prediction is compatible with data at a given confidence level. Finally, we perform a validation of our analysis set-up by reproducing the published bounds on the production cross section of $\Zp$ bosons with given width from the ATLAS collaboration~\cite{Aad:2019fac}.

\subsection{Signal prediction}
\label{sec:Detector}

For a given bin $i$ covering some range of $\mll$, the prediction for the number of detected electron or 
muon pairs ($\ell=e,\mu$) is written as
\begin{equation}\label{eq:Predicted_nbr_of_events}
s_i^\ell = \mathcal{L} \int\d{\mll} \;   \xi_{\ell} (\mll ) W_i \rund{\mll} \frac{\d{\sigma_{\ell}}}{\d{\mll}} \; ,
\end{equation}
where $\mathcal{L}$ is the luminosity, $\frac{\d{\sigma_\ell}}{\d{\mll}}$ is the differential signal cross section  
including interference, 
$W_i \rund{\mll}$ denotes a window function reflecting the finite detector resolution, and $\xi_{\ell} (\mll ) $ is a rescaling factor
taking into account higher-order corrections and detector efficiencies. The different ingredients of the predictions will be
explained in the following.

We perform a fully differential leading-order (LO) computation for the Drell-Yan cross section including a $Z^{\prime}$ mediator. The 
computation is implemented in a fast and efficient computer code as further detailed
in section~\ref{sec:ZPEED} and appendix~\ref{app:xsec}. The fiducial phase-space volume of the ATLAS analysis is defined by $p_T > 30\,$GeV for 
electrons as well as muons. Concerning rapidity, we accept electrons with  $|\eta| < 1.37$ or $1.52 < |\eta| < 2.47$ and
muons with $|\eta| < 2.5$. Integrating over the fiducial volume for fixed invariant dilepton mass $\mll$, 
we obtain $\frac{\d{\sigma_{\ell}}}{\d{\mll}}$.
We also calculate the SM Drell-Yan background $\frac{\d{\sigma^\text{SM}_{\ell}}}{\d{\mll}}$,
i.e.\ the first three terms in \eqn{eq:SigmaHatSplit}, in complete analogy to the signal.

The limited detector resolution is reflected in our analysis using a simple Gaussian kernel which smears the calculated 
invariant mass spectrum. For a bin defined by \mbox{$\mll \in [a_i,b_i]$} the Gaussian smearing is implemented using the window function 
\begin{equation}
W_i(\mll )  = \frac{1}{2} \eck{   \text{erf} \rund{ \frac{b_i - \mll}{ s(\mll ) \sqrt{2}  }  } -   \text{erf} \rund{
\frac{a_i - \mll}{ s(\mll ) \sqrt{2}  }  } } \; ,
\label{eq:Window_function}
\end{equation}
where the detector resolution $s(\mll)$ is taken from the auxiliary figures of the ATLAS analysis in~\citere{Aad:2019fac}.

Unfortunately, detector efficiencies cannot be included at the fully differential level since we lack the full
experimental information. In particular, quality requirements for the muon or electron identification cannot be
approximated by a simple detector simulation like \delphes\ \cite{deFavereau:2013fsa}.\footnote{In particular, \delphes\ significantly overestimates the muon efficiency, which according to Ref.~\cite{Aad:2019fac} should lie between $64\,$\% and $69\,$\%.} 
However, we can make use of the published predictions for the SM Drell-Yan background in order to  estimate detector efficiencies as a function of $\mll$ and then improve our LO prediction $\frac{\d{\sigma^\text{SM}_{\ell}}}{\d{\mll}}$ by appropriate rescaling factors $\xi_{\ell}(\mll)$. In addition, the rescaling also approximately captures higher-order corrections beyond LO in perturbation theory as discussed at the end of the section. 

The rescaling factors $\xi_{\ell}(\mll)$ are derived as follows. Tables 3 and 4 in \citere{Aaboud:2017buh} list the expected event yields $s^\text{exp}_{\ell,i}$ for the Drell-Yan 
background in wide bins of $\mll$.\footnote{Note that the more recent ATLAS analysis \cite{Aad:2019fac} that we use to calculate our final bounds does not provide such information. Since the selection cuts of both analyses are very similar, the rescaling factors obtained in this way can also be applied to the more recent analysis.} We calculate the corresponding event yields 
\begin{equation}
 s^\text{LO}_{\ell,i} =  \mathcal{L} \int\d{\mll} \;  
 W_i \rund{\mll,a_i,b_i} \frac{\d{\sigma^\text{SM}_{\ell}}}{\d{\mll}} \; 
\end{equation}
based on our LO calculation including detector resolution. We then define
$\xi_{\ell}(\mll^i)=s^\text{exp}_{\ell,i}/s^\text{LO}_{\ell,i}$,
where $\mll^i = (a_i + b_i)/2$ is the central $\mll$ value in a given bin $i$ with $\mll \in [a_i,b_i]$. The rescaling factors obtained in this way are stated in \tab{tab.:interpolation_xi}. The function 
$\xi_{\ell}(\mll)$ is then obtained by linear interpolation. Since the resulting functions $\xi_{\ell}(\mll)$ depend only weakly on $\mll$, the simple linear interpolation turns out to be a sufficient approximation.

\begin{table}[t]
\begin{center}
  \begin{tabular}{ | c || c | c | c | c | c | c | c | c | c | c | c | c | }
    \hline
    $\mll^i\,\lbrack$GeV$\rbrack$ & 80 & 100 & 185 & 325 & 450 & 600 & 800 & 1050 & 1500 & 2400 & 4500 \\ \hline\hline
    $\xi_e$ & 0 & 0.71 & 0.88 & 1.06 & 1.11 & 1.11 & 1.09 & 1.08 & 1.06 & 0.97 & 0.87\\ \hline
    $\xi_{\mu}$ & 0 & 0.56 & 0.63 & 0.65 & 0.65 & 0.63 & 0.59 & 0.59 & 0.55 & 0.50 & 0.51 \\ \hline
  \end{tabular}
  \end{center}
  \vspace{-2mm}
  \caption{Interpolation nodes and values of $\xi_{\ell}$. For higher values of $\mll$, we do not extrapolate but take the maximum value given in the table as efficiency rescaling function.}\label{tab.:interpolation_xi}
  \vspace{-2mm}
  \end{table}

As an alternative approach, we have first used \delphes\ on a fully differential level to include those detector effects that are implemented.
Additional detector effects not included in \delphes\ are then again included by our rescaling approach. The differences between the two approaches are negligible. Hence, for simplicity, we do not use any detector simulation by \delphes\ for the results presented in the following.

As noted above, our LO cross section is not only modified by detector effects but also by higher-order corrections. 
The dominant higher-order corrections are approximately included in our rescaling procedure as well because they are included in the expected event yields $s^\text{exp}_{\ell,i}$. Like the detector efficiency, the corrections are not included at the fully differential level but they are effectively treated as $\mll$-dependent K-factors along with the detector effects. Here, we assume that the higher-order corrections
affect the SM background in the same way as the differential signal cross section including interference. This is certainly true for the QCD corrections to the Drell-Yan process, which only concern the initial state. 

\subsection{Statistical method}
\label{sec:Statistics}

Having calculated the predicted signal $s_i^\ell$ in each bin, we can construct the likelihood
\begin{equation}
 -2 \log \mathcal{L}(\mu) = 2 \sum_{\ell = e, \mu} \sum_i \mu s_i^\ell + b_i^\ell - o_i^\ell + o_i^\ell \log \left( \frac{o_i^\ell}{\mu s_i^\ell + b_i^\ell}\right) \; ,
 \label{eq:likelihood}
\end{equation}
where $b_i^\ell$ and $o_i^\ell$ denote the expected background and the observed number of events, respectively, and we have introduced the signal strength modifier $\mu$. The background estimates $b_i^\ell$ may depend on additional nuisance parameters, in which case $-2 \log \mathcal{L}(\mu)$ denotes the profile likelihood (where all nuisance parameters have been set to the values that maximise the likelihood for given $\mu$). The contribution from interference between signal and background is included in the predicted signal $s_i^\ell$. 
Since signal and interference depend differently on the parameters of the underlying model, the term $\mu s_i^\ell$ is unphysical for general values of $\mu$ in the sense that it does not correspond to any parameter combination. Nevertheless, introducing $\mu$ is a useful construction to interpolate between the signal+background hypothesis ($\mu = 1$) and the background-only hypothesis ($\mu = 0$) without changing the shape of the signal.

The value of $\mu$ that maximises the likelihood is called $\hat{\mu}$. Having found this value, we calculate the test statistic
\begin{equation}
 q_\mu = -2 (\log \mathcal{L}(\mu = 1) - \log \mathcal{L}(\hat{\mu})) \; ,
\end{equation}
which is expected to follow a $\chi^2$ distribution with 1 degree of freedom. Rather than calculating exclusion bounds directly from $q_\mu$, we employ the $\mathrm{CL_s}$ method~\cite{ATLAS:2011tau}. In the asymptotic regime ($b_i, o_i \gg 1$), the modified $p$-value of the signal+background hypothesis is given by\footnote{For large values of $m_{Z'}$ the assumption of asymptotics leads to exclusion limits that are too strong by a factor of 2 or more. The main focus of the present work is however on $m_{Z'} \lesssim 2 \, \mathrm{TeV}$, where the asymptotic expression for $\mathrm{CL_s}$ provides a very good approximation.}
\begin{equation}
 \mathrm{CL_s} = \frac{1 - \Phi(\sqrt{q_\mu})}{\Phi(\sqrt{q_{\mathrm{A},\mu}} - \sqrt{q_\mu})} \; .
\end{equation}
Here $\Phi$ is the cumulative distribution function of the normal distribution and $q_{\mathrm{A},\mu}$ is the value of the test statistic $q_\mu$ for the Asimov data set~\cite{Cowan:2010js}, in which all observations exactly match the background expectation ($o_i^\ell = b_i^\ell$), such that $\hat{\mu}_\mathrm{A} = 0$.

The signal+background hypothesis ($\mu = 1$) can now be rejected with (at least) $95\,$\% confidence level if $\mathrm{CL_s} \leq 0.05$. It is common practice to solve $\mathrm{CL_s} = 0.05$ for $\mu$ in order to find the smallest value of $\mu$ that is excluded. However, as discussed above only $\mu = 0$ and $\mu = 1$ represent actual physical models. 
In the following, we will therefore not quote bounds on $\mu$ but instead apply the $\mathrm{CL_s}$ method to every point in parameter space in order to identify those parameter regions where $\mu = 1$ is excluded.

At present only ATLAS provides publicly available data for dilepton resonance searches based on the entire data from Run 2~\cite{Aad:2019fac}, and we focus on their analysis here.\footnote{We have checked that including the publicly available data from CMS based on an integrated luminosity of $36 \, \mathrm{fb^{-1}}$~\cite{Sirunyan:2018exx} does not substantially change any of the results that we present.} In contrast to previous dilepton resonance searches, ATLAS does not rely on Monte Carlo simulations to estimate backgrounds, but instead obtains the background estimates by fitting a smooth function to the observed data. In principle, the uncertainties on the fit parameters obtained in this way should be included as nuisance parameters. 
However, given that the background is fitted across many different bins, while the signal is more localised, the uncertainties in the nuisance parameters have a negligible impact on the profile likelihood. For our implementation we therefore simply take $b_i^\ell$ to be the central value of the background prediction.

To reproduce the ATLAS analysis as closely as possible, we exclude the contribution from off-shell $Z'$ bosons at small $m_{\ell\ell}$. Specifically, we limit ourselves to the signal region defined by $m_{\ell\ell} > m_{\ell\ell,\text{min}} \equiv m_{Z'} - 2 \Gamma_\text{eff}$ with $\Gamma_\text{eff}^2 = \gz^2 + s(m_{Z'})^2$, where $s(m_{\ell\ell})$ is the detector resolution in the $e^+ e^-$ channel. In general $m_{\ell\ell,\text{min}}$ will not coincide with the boundary of any bin. The bin $[a,b]$ that satisfies $a < m_{\ell\ell,\text{min}} < b$ is included in the likelihood, but its contribution is multiplied with the weighting factor
\begin{equation}
 w_i = \frac{s_\text{min}}{s_i} \; ,
\end{equation}
where $s_\text{min}$ is the number of signal events in the interval $[m_{\ell\ell,\text{min}}, b]$. This approach ensures that the likelihood is a continuous function of $m_{Z'}$. 
We impose no upper bound on $\mll$ other than the requirement $\mll < 6253\,\mathrm{GeV}$ implied by the ATLAS data.

\subsection{Validation}
\label{sec:Validation}

In order to check our implementation of the detector efficiencies, smearing and rescaling functions, as well as our statistical analysis, in this subsection we validate our results by comparing to bounds published by the ATLAS collaboration in \citere{Aad:2019fac}.

\begin{figure}[t]
    \centering
        \includegraphics[width=0.49\textwidth]{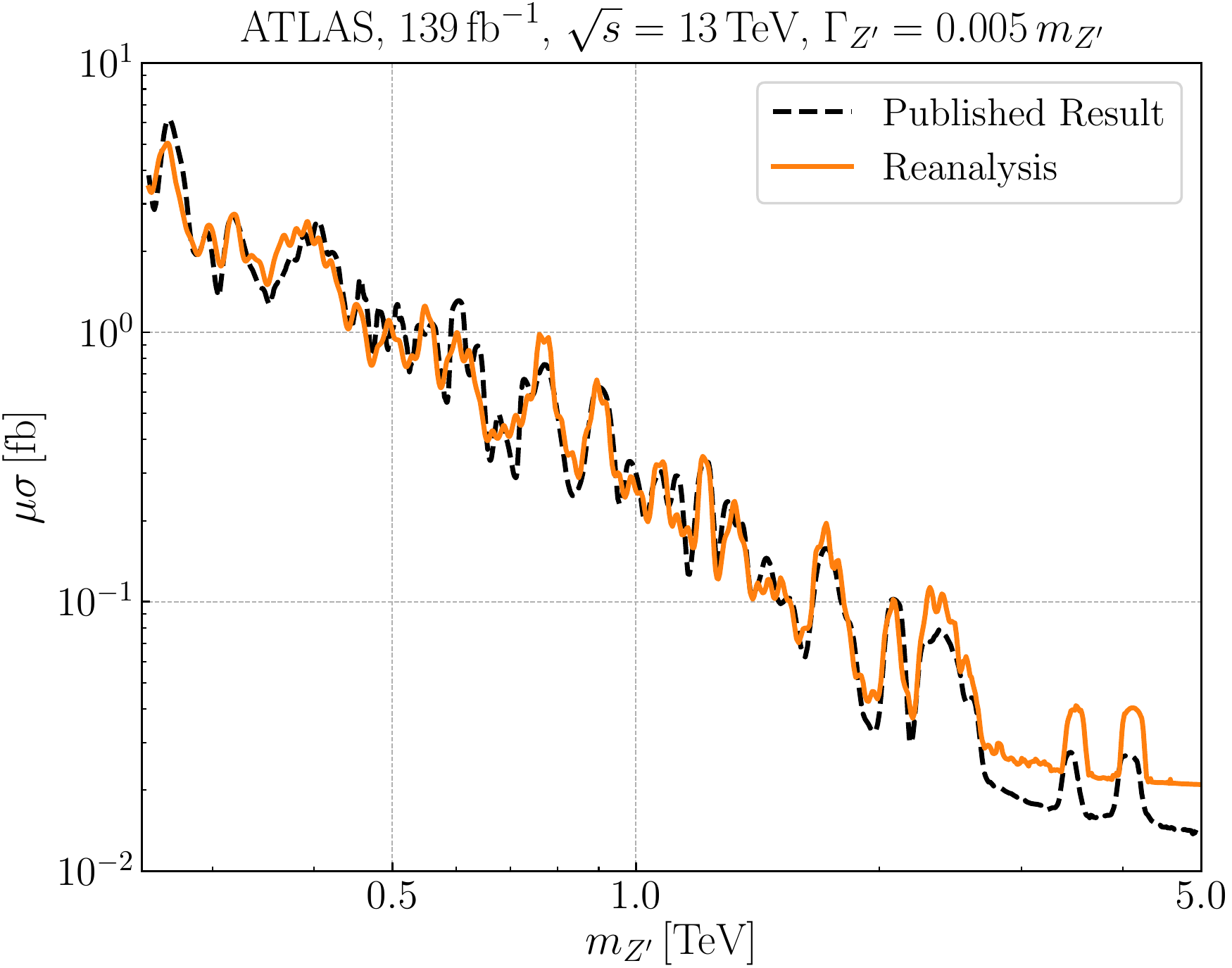}
        \includegraphics[width=0.49\textwidth]{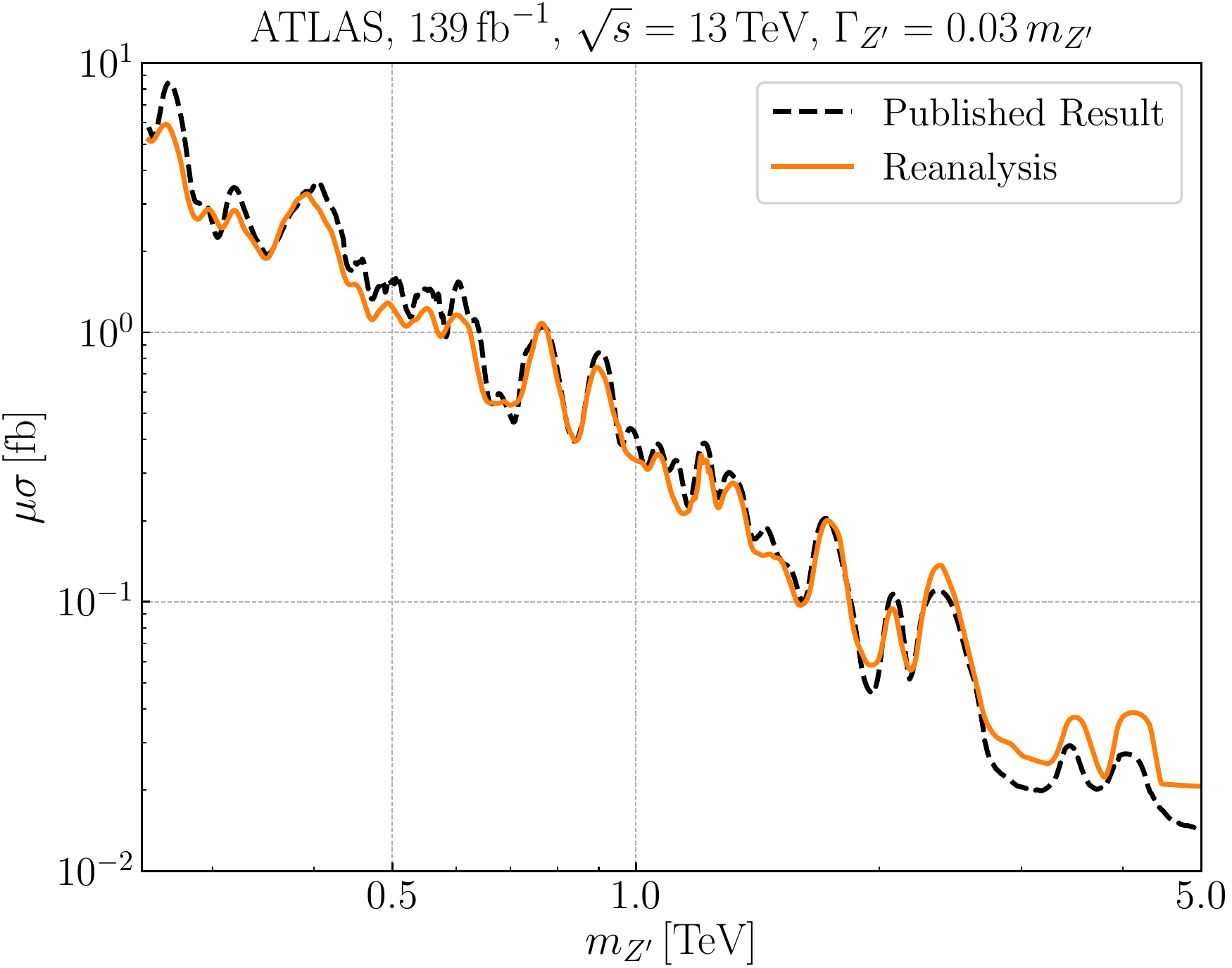}
        \vspace{-2mm}
    \caption{$95\,$\% confidence limits on signal strength $\mu$ times predicted signal cross section $\sigma$ for a $Z^\prime$ decaying into $e^+ e^-$ or $\mu^+ \mu^-$ at ATLAS with $\gz = 0.005\,  m_{Z^\prime}$ (left) and with $\gz =  0.03 \, m_{Z^\prime}$ (right). The axial couplings are set to zero and interference terms are neglected. The experimental bounds are taken from \citere{Aad:2019fac}.}
    \vspace{-2mm}
\label{fig:Validation}
\end{figure}

We show in \fig{fig:Validation} our bound on the cross section as a function of the $\Zp$ mass, compared to that published by ATLAS~\cite{Aad:2019fac}. Note that in order to reproduce the approach taken by the experimental analysis, these bounds are calculated ignoring the effect of interference with SM Drell-Yan processes. The bounds are for $95\,$\%~C.L.\ and show good agreement for a $\Zp$ width of $0.5\,\%$ (left) and $3\,\%$ (right). We have also checked the bounds for larger $\Zp$ widths and find good agreement up to $\gz / \mz \approx 6\%$. For even larger widths correlated background uncertainties, which cannot be properly included with publicly available information, become important and our approach yields bounds that are slightly stronger than the published ones. For signal widths smaller than $0.5\,\%$, on the other hand, bounds will be dominated by detector resolution and will be very similar to the case shown in the left panel of \fig{fig:Validation}. We hence conclude that our implementation is reliable for any $\Zp$ signals with $\gz/\mz \lesssim  0.06$.

\subsection{ZPEED}
\label{sec:ZPEED}

To obtain these results, we have developed a highly efficient numerical code called \texttt{ZPEED} (\emph{$\Zp$ Exclusions from Experimental Data}), which is capable of calculating the likelihood and $\mathrm{CL_s}$ value for a given $\Zp$ parameter point within less than a second on a single CPU. The code implements the approach outlined in appendix~\ref{app:xsec}, i.e.\ it uses analytical expressions for the differential cross sections of signal and interference terms together with tabulated values of the function $\Tf{}{q,2}(\mll)$ as defined in eq.~(\ref{eq.:def_Tf}), which accounts for PDFs and phase space cuts. The differential cross sections are then multiplied with the rescaling factors $\xi_\ell(\mll)$ and the window functions $W_i(\mll)$ introduced in eq.~(\ref{eq:Predicted_nbr_of_events}). Indeed, the integration over $\mll$ in eq.~(\ref{eq:Predicted_nbr_of_events}), which needs to be performed at runtime, is the only computationally expensive step. Once the predictions $s_i^\ell$ have been calculated, it is straight-forward to calculate the likelihood defined in eq.~(\ref{eq:likelihood}) as a function of the signal strength $\mu$, determine $\hat{\mu}$ and obtain the $\mathrm{CL_s}$ value. At present only the ATLAS analysis based on $139\,\mathrm{fb^{-1}}$ has been implemented, but future updates will be provided whenever new data becomes publicly available. The code is open source and can be downloaded from \url{https://github.com/kahlhoefer/ZPEED}.

\section{Results}
\label{sec:DM}

In this section we illustrate the importance of interference effects by showing how they impact bounds derived from experimental data. We will first do this in a model-independent way by treating couplings and width as independent parameters and then focus on a specific simplified model, in which the width of the $Z'$ is calculated self-consistently as a function of the underlying parameters. 

\subsection{Model-independent bounds}

We first consider a general $Z'$ model with vector couplings and define the effective coupling $g \equiv (\gqv \glv)^{1/2}$. For fixed total width $\gz$ the $Z'$ production cross section is proportional to $g^4$, while interference effects scale as $g^2$. We can therefore use the analysis chain presented in section~\ref{sec:Analysis} to calculate bounds on $g$ with and without interference for different values of $\gz$.

\begin{figure}[t]
	\centering    
        \includegraphics[width=0.58\textwidth]{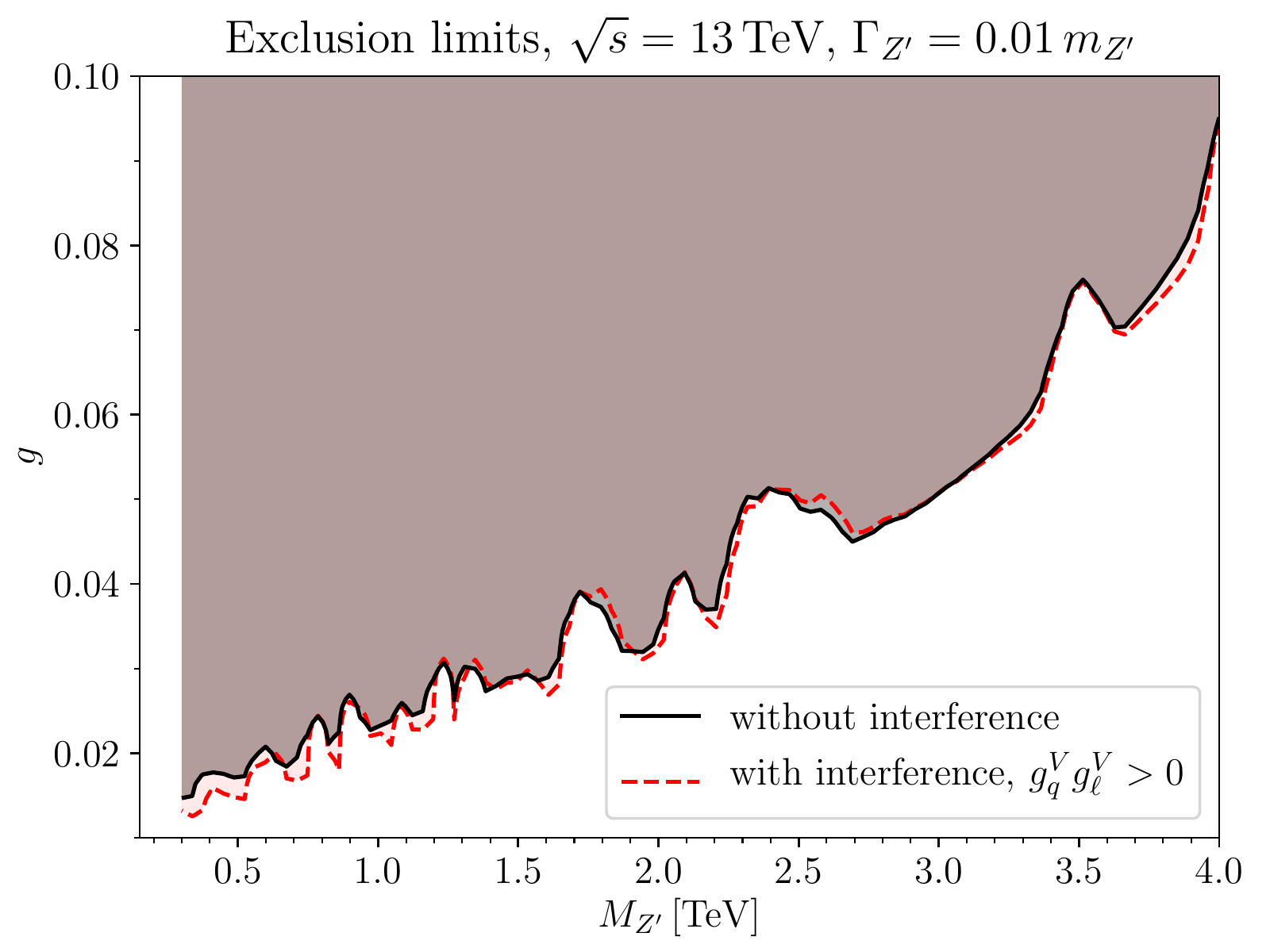} 
        \includegraphics[width=0.58\textwidth]{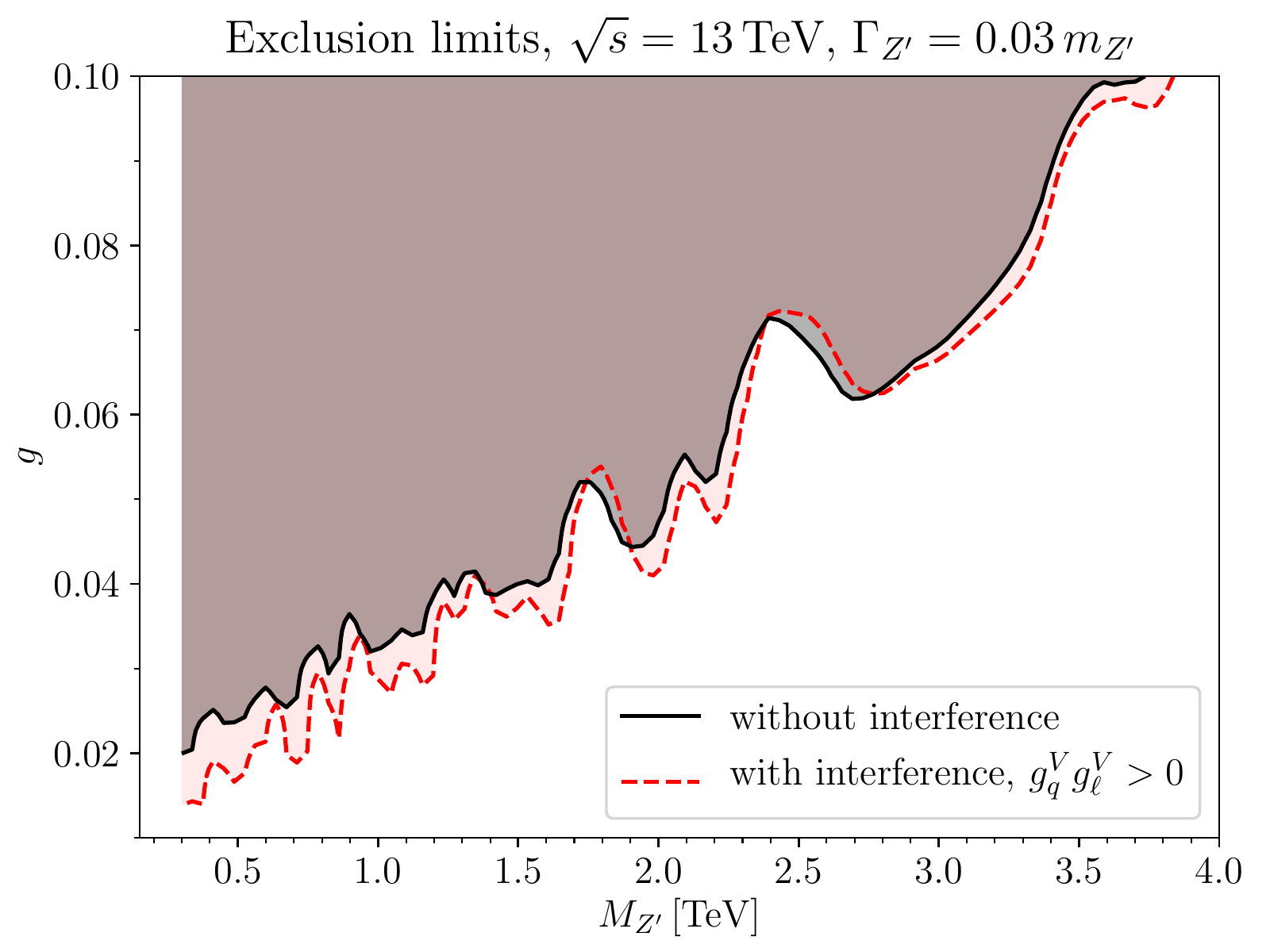}
        \includegraphics[width=0.58\textwidth]{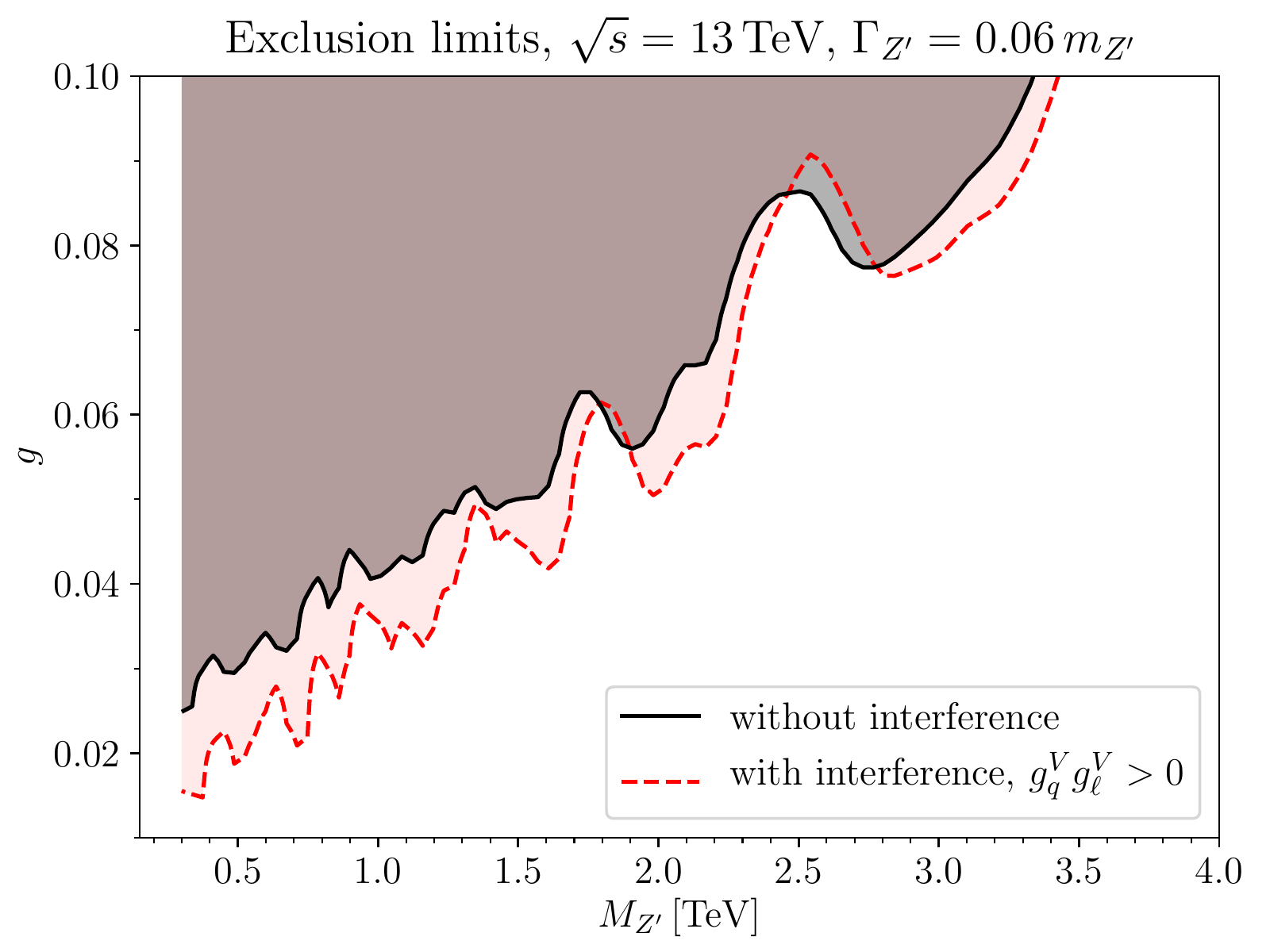}
    \caption{Upper bound on the effective coupling $g = (\gqv \, \glv)^{1/2}$ at $95\,$\% confidence level, with and without interference effects. We consider $\Zp$ bosons with vanishing axial couplings and different relative widths $\gz / \mz = $ 1\%, 3\%, 6\%.
    }
    \vspace{-2mm}
    \label{fig:chi2_couplings}
    \vspace{-2mm}
\end{figure}

The resulting exclusion bounds are shown in figure \ref{fig:chi2_couplings} as a function of the mediator mass $\mz$ for $\gz/\mz = 0.01, 0.03$ and $0.06$. As expected, interference effects are negligible when the relative width is small (top panel) and become increasingly important as $\gz/\mz$ increases. Interference effects are largest for small values of $\mz$, which is a consequence of the steeply falling SM background. In the bottom panel, which assumes a $6\,\%$ relative width, interference effects lead to a strong enough distortion of the input signal such that the exclusion limits are changed significantly. For instance, for $\mz \approx 500\,$GeV, the exclusion limit on $g$ obtained from the pure $\Zp$ signal is about a factor of 1.5 weaker if interference effects are neglected. Moreover, interference shifts the position of the peak in the dilepton invariant mass spectrum to smaller values (see figure \ref{fig:InterferenceSignals}), which results in a shift of the exclusion bound to larger masses. For example, the dip around $\mz \approx 1.9\,\mathrm{TeV}$ in the bottom panel is shifted to about $\mz \approx 2\,\mathrm{TeV}$ once interference effects are included.

Instead of assuming equal couplings to electrons and muons, one can also calculate constraints on the effective coupling $g$ to each lepton family separately. The resulting upper bounds are provided in appendix~\ref{app:opposite}.

We emphasize that for large relative widths the impact of interference effects is at least as important as the impact of higher-order QCD corrections. In particular, the former can significantly change the shape of the signal, while the latter only result in an effective rescaling of the cross section that can be applied after signal events have been generated. Interference effects, on the other hand, need to be included during signal generation and depend in a more complicated way on the underlying parameters. It is essential to include these effects in order to obtain accurate bounds on the parameter space of a given $\Zp$ model. In most cases including interference effects leads to stronger exclusion limits, which further enhances the potential of dilepton resonance searches to constrain models of BSM physics. 

\subsection{Bounds on dark matter simplified models}

As we have seen above, interference effects are most important for large relative widths. Such large widths typically cannot be obtained from decays into SM particles (as the required couplings would violate experimental constraints), but they are a generic prediction in models with additional contributions to the $\Zp$ width arising from decays into new invisible light degrees of freedom. As a specific example of such a model, we consider a spin-one simplified DM model~\cite{Abdallah:2015ter}, which has been employed by the LHC collaborations~\cite{Boveia:2016mrp,Albert:2017onk} to create benchmark points in theory space that allow for different LHC DM searches to be compared to each other and to non-collider experiments. 

We extend \eqn{eq:Lagrangian} to include a coupling to a SM singlet Dirac fermion $\chi$ with mass $m_\chi$ as a DM candidate. The corresponding interaction Lagrangian reads
\begin{equation}
\mathcal{L}_\text{int} = - Z^{\prime\mu} \, \bar{\chi} \, \left( g_\chi^V \gamma_{\mu} + g_\chi^A \gamma_{\mu} \gamma^5 \right) \, \chi -  \sum_{f = q, \ell, \nu} \, Z^{\prime\mu} \, \bar{f} \, \left( g_f^V \gamma_{\mu} + g_f^A \gamma_{\mu} \gamma^5 \right) \, f  \; .
\label{eq:LagrangianV}
\end{equation}
Then each partial width of the $\Zp$ is
\begin{equation}\label{eq:DecayWidth}
\Gamma\rund{Z^{\prime}\longrightarrow f\bar{f}} = \frac{\mz N_c}{12\pi } \sqrt{1-\frac{4m_f^2}{\mz^2}}  \left( \rund{\gfv}^2  +  \rund{\gfa}^2  + \frac{m_f^2}{\mz^2} \left( 2 \rund{\gfv}^2 - 4 \rund{\gfa}^2 \right) \right) \; ,
\end{equation}
where $N_c$ is the number of colours. 

It has been shown that (for a minimal Higgs sector) gauge invariance requires $g^A_l = g^A_q$~\cite{Kahlhoefer:2015bea}, which typically leads to overwhelmingly strong constraints from dilepton resonance searches in models with non-zero axial couplings. We therefore focus on the case $g^A_{q/ \ell /\chi} = 0$, while the three remaining couplings $\gqv$, $\glv$ and $\gxv$ are treated as independent parameters. A particularly well-motivated possibility is that $\glv$ vanishes at high scales and is only introduced at low scales through kinetic mixing~\cite{Duerr:2016tmh}. In this case one naturally finds $\gqv \gg \glv > 0$, such that bounds from dilepton resonance searches are suppressed but still relevant.

In the simplified DM model introduced above, $\gz$ depends decisively on the mass hierarchy. For $m_\chi > \mz / 2$, invisible decays are kinematically forbidden and the relative width is very small. In the opposite case, the partial width $\Gamma\rund{Z^{\prime}\longrightarrow \chi\bar{\chi}}$ may contribute significantly to the total width, in particular if $\gxv \gg \gqv$. Following the recommendations of the LHC DM working group~\cite{Albert:2017onk}, we therefore consider the benchmark choice $\gxv = 1.0$ and $\gqv = 0.1$, such that $\gz / \mz \approx 0.5\,\%$ for $m_\chi > \mz / 2$ and $\gz / \mz \approx 3\,\%$ for $m_\chi \ll \mz / 2$. We consider the two choices $\glv = 0.01,\, 0.02$, corresponding to an effective coupling $g = (\gqv \, \glv)^{1/2} = 0.032$ and $g = 0.045$, respectively. 

\begin{figure}[t]
    \centering
    \includegraphics[width= 0.59 \textwidth]{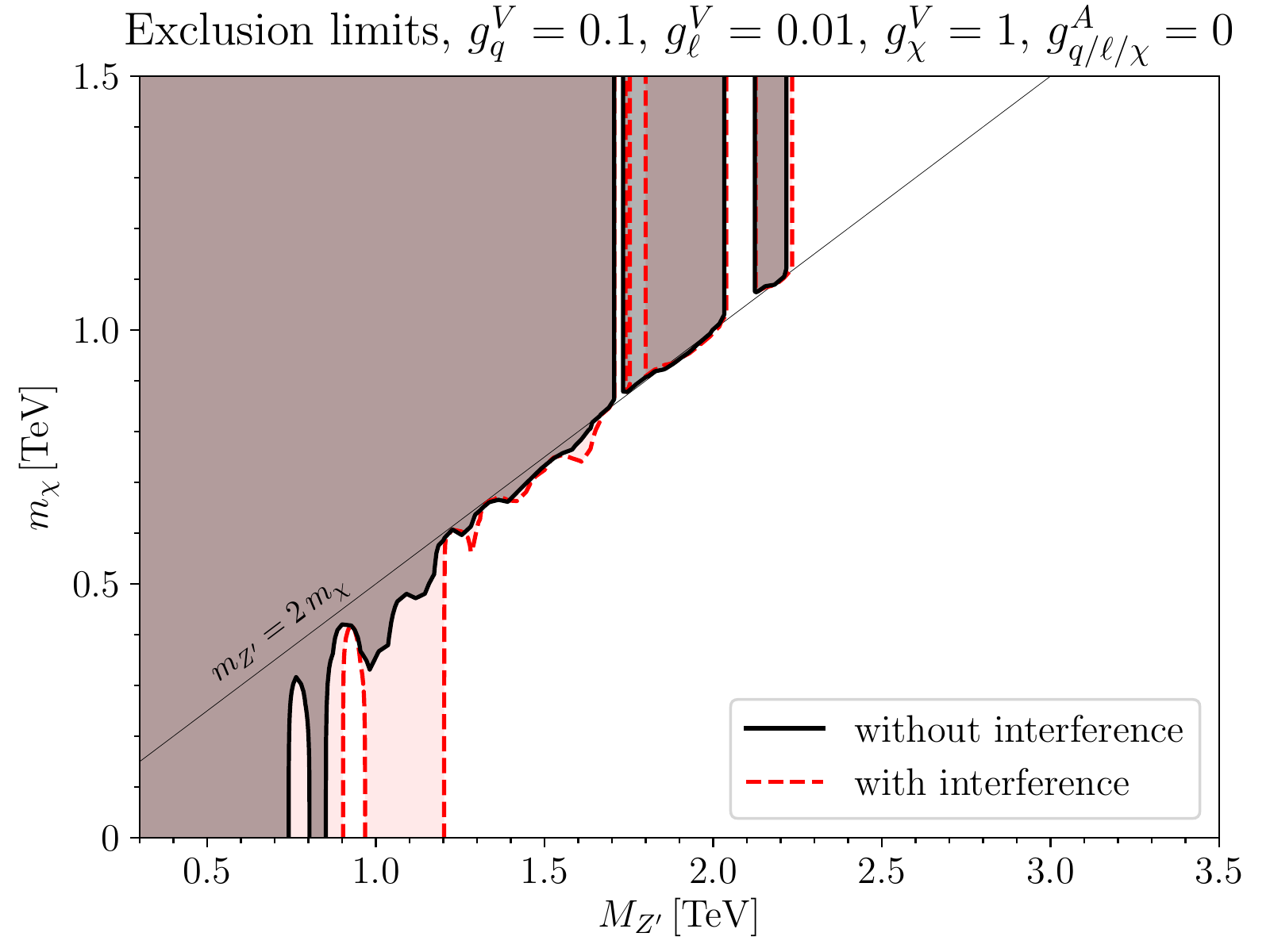} 
    \includegraphics[width= 0.59 \textwidth]{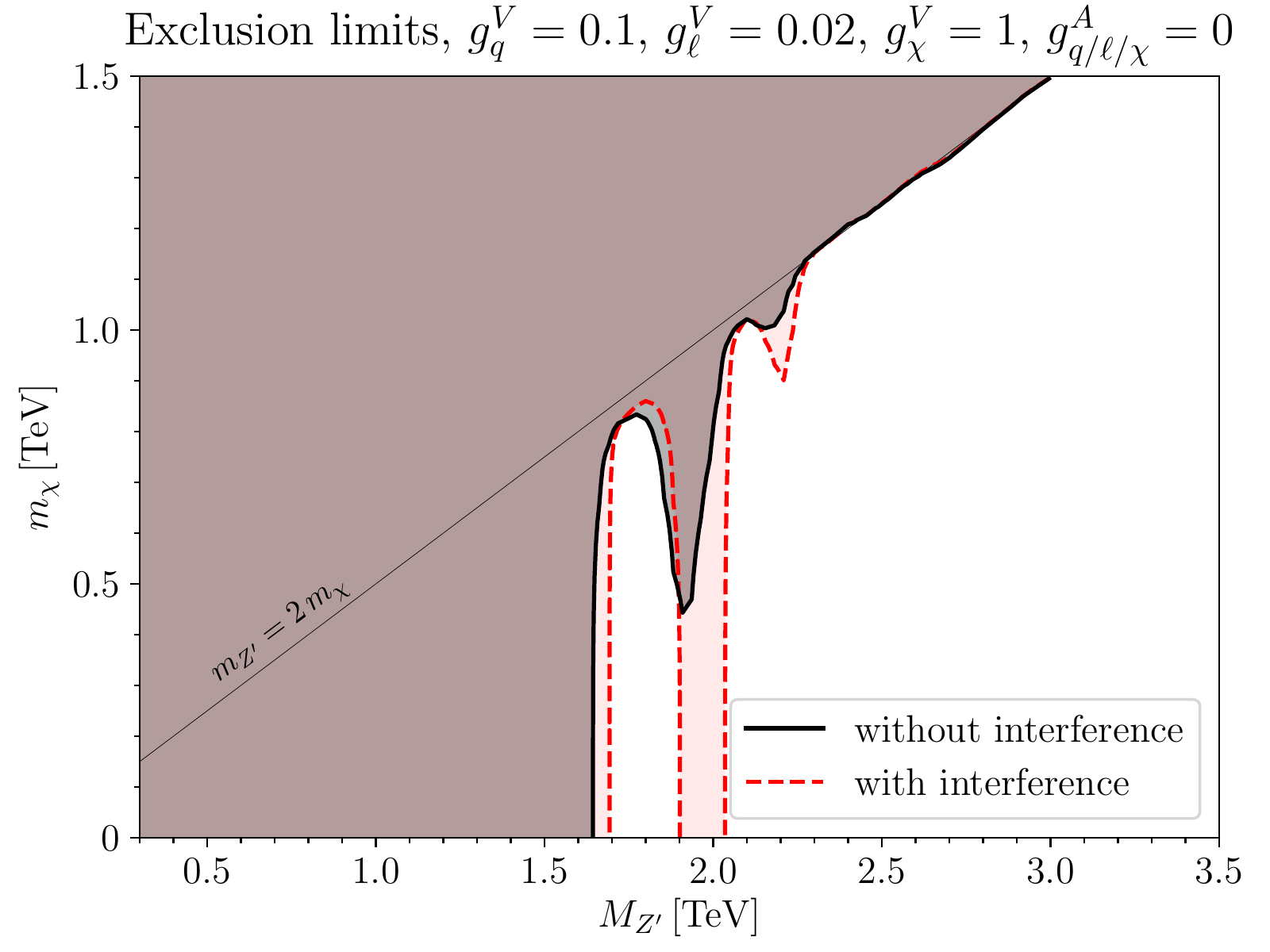}
    \vspace{-2mm}
    \caption{Excluded parameter space at $95\,$\% confidence level in the $\mz$-$m_\chi$ parameter plane with and without interference effects for a simplified DM model with vanishing axial couplings. Both panels assume $\gxv = 1.0$ and $\gqv = 0.1$, while the lepton coupling is set to $\glv = 0.01$ in the top panel and $\glv = 0.02$ in the bottom panel.     }\label{fig:chi2_mass}
    \vspace{-2mm}
\end{figure}

Figure~\ref{fig:chi2_mass} displays the resulting exclusion limits in the $\mz$-$\mx$-plane both with and without the inclusion of interference effects.\footnote{We note that the exclusion limit obtained in the absence of interference effects is slightly stronger than the one provided by the ATLAS collaboration. In the absence of a detailed documentation it is difficult to identify the origin of this discrepancy.} We emphasize again that in these plots the decay width is computed following \eqn{eq:DecayWidth}. As expected, interference effects are most important for $m_\chi < \mz / 2$, corresponding to larger relative width of the $\Zp$, and for small $\glv$. When interference effects are neglected, the parameter region with small $m_\chi$ is essentially unconstrained for $\mz \gtrsim 850 \, \mathrm{GeV}$ ($\mz \gtrsim 1650 \, \mathrm{GeV}$) in the case that $\glv = 0.01$ ($\glv = 0.02$). Including interference effects, the parameter region probed by dilepton resonance searches is extended to $\mz \lesssim 1200 \, \mathrm{GeV}$ ($\mz \lesssim 2050 \, \mathrm{GeV}$). Although the precise parameter regions excluded by the ATLAS analysis depend sensitively on fluctuations in the data, the general trend is clear: interference effects lead to stronger bounds on the simplified DM model.

To conclude this discussion, we note that $\chi$ should, as a DM candidate, also satisfy bounds coming from the relic density of DM and from direct and indirect detection experiments in addition to collider bounds. A number of works have investigated in detail the complementarity of these different constraints (see e.g.~\citeres{Malik:2014ggr,Abdallah:2015ter,Bauer:2016gys,Boveia:2016mrp,Albert:2017onk,Frandsen:2012rk,Fox:2012ru,Alves:2013tqa,Arcadi:2013qia,Buchmueller:2014yoa,Lebedev:2014bba,Harris:2014hga,Busoni:2014gta,Fairbairn:2014aqa,Jacques:2015zha,Alves:2015pea,Chala:2015ama,DeSimone:2016fbz,Brennan:2016xjh,Jacques:2016dqz,Blanco:2019hah,Kahlhoefer:2015bea,Duerr:2016tmh,Duerr:2017uap,Ellis:2018xal,Caron:2018yzp,ElHedri:2018cdm}). Here we focus on the contribution of $\chi$ to the total decay width of the $\Zp$ and the resulting interference effects. Therefore, we do not make any assumptions on the cosmological history and the relic abundance of $\chi$. In fact, all of the results presented in this work remain valid even if $\chi$ is unstable and decays into either SM particles or other BSM states. 

\section{Conclusions}\label{sec:Conclusion}

In this work we have investigated the sensitivity of the LHC to new $\Zp$ bosons with a focus on the effect of interference between the $\Zp$ signal and SM Drell-Yan background in the dilepton channel. Interference is enhanced for $\Zp$ bosons with large width (compared to the detector resolution), arising for example from invisible decay modes into new light degrees of freedom, and results in an asymmetric signal with modified peak amplitude and position (see figure \ref{fig:InterferenceSignals}). Details of our calculations and of the fast numerical implementation can be found in appendix~\ref{app:xsec}. 

In order to quantify the impact of interference on bounds derived from experimental data, we have implemented an existing ATLAS search for dilepton resonances. We use smearing functions to model energy resolution based on experimental data and estimate experimental efficiencies and higher-order corrections by rescaling our predicted Drell-Yan background to published background estimates. We have calculated exclusion bounds on the fiducial cross section neglecting interference with the $\mathrm{CL_s}$ method and found excellent agreement with published limits (see figure \ref{fig:Validation}). We have made the code used to obtain these results publicly available.\footnote{\texttt{ZPEED} -- \emph{$Z'$ Exclusions from Experimental Data}: \url{https://github.com/kahlhoefer/ZPEED}.}

We then applied this analysis to the case of a $\Zp$ with purely vectorial couplings in order to obtain bounds on the effective coupling $g= (g_q g_\ell)^{1/2}$ as a function of $m_{\Zp}$ for different values of $\gz$ (see figure \ref{fig:chi2_couplings}). As expected, interference effects are most important for large widths and can substantially strengthen the bounds on the effective coupling $g$. For example, for a $\Zp$ with $6\,$\% relative width the bound on the couplings improves by up to a factor of 1.5 once interference is included.

We also considered a specific example for a model where the $\Zp$ width can be large in spite of small couplings to quarks and leptons, namely a simplified model of DM with a spin-1 mediator. Assuming  the DM coupling $g_\chi$ is large compared to $g_q$ and $g_\ell$, decays of the mediator into DM particles give rise to a large invisible width and therefore a substantial increase of the total width, whenever decays into DM are kinematically allowed. In this model the $\Zp$ width can easily be large compared to the detector resolution and therefore large enough for interference effects to be relevant. As a specific benchmark we considered $g_\chi = 1$, $g_q = 0.1$, and $g_\ell = 0.01$ (a choice recommended by the LHC Dark Matter Working Group and used by both ATLAS and CMS to present exclusion limits~\cite{Albert:2017onk}), as well as an additional example with $g_\ell = 0.02$. Deriving bounds on this model as a function of DM mass and mediator mass, we demonstrated that interference effects lead to substantially stronger constraints on the parameter space of this model (see figure \ref{fig:chi2_mass}). 

The LHC is entering the phase where precise signal predictions are essential in order to fully exploit the benefits of high statistics. We argue that in order to accurately calculate constraints on $\Zp$ bosons with large widths from dilepton resonances, interference effects must be included. This is particularly true in DM models where the $\Zp$ acts as the mediator between DM and SM fermions and obtains a large invisible width. We encourage the experimental collaborations to incorporate the modified signal shapes and look forward to the inclusion of interference effects in bounds coming from existing and new LHC data.

\acknowledgments

We thank Caterina Doglioni, Ulrich Haisch, Michael Kr{\"a}mer and Susanne Westhoff for discussions. This  work  is  funded  by  the  Deutsche Forschungsgemeinschaft (DFG) through the Collaborative Research Center TRR 257 ``Particle Physics Phenomenology after the Higgs Discovery'', the Emmy Noether Grant No.\ KA 4662/1-1 and the Research Unit FOR 2239 ``New Physics at the Large Hadron Collider''.

\appendix

\section{Cross section calculations}
\label{app:xsec}

In this appendix, we provide more details on the calculation of the partonic and hadronic
Drell-Yan cross section in the $\Zp$ model under consideration.
The interaction Lagrangian of the $\Zp$ has been introduced in \eqn{eq:Lagrangian}. 
The differential LO result for the partonic signal cross section $\hat{\sigma}_{\Zp\Zp}$ reads
\begin{equation}\label{dsigmadt_for_final_xsec}
\frac{\d{\hat{\sigma}_{\Zp\Zp}}}{\d{\hat{t}}} =   \frac{1}{8\pi N_c} \frac{1}{\rund{\ps-\mz^2}^2+\mz^2 \gz^2}
 \eck{c_0^q + c_1^q \cdot \frac{\hat{t}}{\ps}  + c_2^q \cdot  \frac{\hat{t}^2}{\ps^2}} \, ,
\end{equation}
where $\hat{s}$ and $\hat{t}$ are the usual Mandelstam variables, $N_c=3$ for QCD, $\mz$ is the
mass of the $\Zp$ and $\gz$ its total width. The coupling coefficients read
\begin{equation}
\begin{split}
c_0^q &= \eck{ \rund{\gqv}^2 + \rund{\gqa}^2  } \cdot  \eck{ \rund{\glv}^2 + \rund{\gla}^2  } - 4\gqv \gqa \glv \gla \; ,\\
c_1^q &=  2 c_0^q \; , \\
\text{and}   \;\;\;\;\;\; c_2^q &= 2 \eck{ \rund{\gqv}^2 + \rund{\gqa}^2  } \cdot  \eck{ \rund{\glv}^2 + \rund{\gla}^2  } \; .
\end{split}
\end{equation}
Convolving the partonic cross section with parton-distribution functions of the quarks $f_{q}$ and anti-quarks
$f_{\bar{q}}$, the fully differential 
hadronic cross section is given by

\begin{equation}\label{LHC_Code_Formel}
\frac{\d{}^3 \sigma_{\Zp\Zp}}{\d{\eta_+}\d{\eta_-}\d{\mll}} = \frac{1}{2} \sum_q x_1 f_q (x_1)\, x_2 f_{\bar{q}} (x_2)
\; \frac{\mll}{\cosh^2 y}\, \frac{\d{\hat{\sigma}_{\Zp\Zp}}}{\d{\hat{t}}} \; ,
\end{equation}
where $\mll=\sqrt{\hat{s}}$ is the dilepton invariant mass, $\eta_\pm$ are the rapidities of the positively and negatively
charged leptons in the lab frame, $x_i$ are the momentum fractions of the partons, and
$y = \frac{1}{2} \rund{\eta_+ -\eta_-}$. The sum runs over all light quark and anti-quark flavours.
To obtain this result, we have made use of the following relations between the different kinematic variables:
\begin{align}
\hat{t} = -\frac{\mll^2}{2 \cosh y} \, e^{-y} \, , & & x_1 = \frac{\mll}{\sqrt{s}} e^{Y}	\,, & & x_2 = \frac{\mll}{\sqrt{s}} e^{-Y} \; ,
\end{align}
where $Y = \frac{1}{2} \left( \eta_+ + \eta_- \right)$.
Hence, we can define
\begin{equation}\label{eq.:def_Tf}
\Tf{}{{q,i}} (\mll) := \int  \d{\eta_+}\d{\eta_-} \;  x_1 f_q (x_1)\, x_2 f_{\bar{q}} (x_2)  \; \frac{1}{\cosh^2 y}\; \rund{ \frac{\hat{t}}{\ps}}^i \; ,
\end{equation}
where $i=0,1,2$ and it is understood that we only integrate over the fiducial region, i.e.\ the cuts
on the rapidities and the lepton transverse momenta $p_T = \mll/ (2 \cosh y)$  are respected. With this definition, we write
the differential cross section
\begin{equation}\label{From_Partons_To_Hadrons_No_Int}
\frac{\d{ \sigma_{\Zp\Zp}}}{\d{\mll}}  =  \frac{1}{16\pi N_c} \frac{\sqrt{\ps}}{\rund{\ps-\mz^2}^2+\mz^2 \gz^2}  
\sum_{i=0}^2 \sum_q  c_i^q \cdot  \Tf{}{q,i}(\mll) 
\end{equation}
as a product of model-independent but cut-dependent function $\Tf{}{q,i}(\mll)$ and simple model-dependent
factors consisting of couplings and propagators.
Employing \texttt{MSTW} parton distribution functions \cite{Martin:2009iq}, 
the $\Tf{}{q,i}$-functions can be evaluated once on a fine discrete $\mll$-grid and 
linearly interpolated, such that no numerical integrations have to be performed when the cross section is
evaluated for different model parameters. Hence, \eqn{From_Partons_To_Hadrons_No_Int} 
is a particularly efficient implementation for parameter scans. Note that 
this separation only works for $s$-channel mediated interactions like the Drell-Yan like process under
consideration, since only in this case the propagator does not depend on the rapidities.

The evaluation of the hadronic cross section further simplifies, since for our (symmetric)
fiducial volume one has
\begin{equation}\label{T0_T1_identity}
\Tf{}{q,0} + 2 \Tf{}{q,1} = 0 \; 
\end{equation}
and $c_1^q = 2 c_0^q$ implies that $\Tf{}{q,0}$ and $\Tf{}{q,1}$ do not contribute to the
cross section.

So far, we have only discussed the signal cross section $\d{ \sigma_{\Zp\Zp}}/\d{\mll}$ without
interference. However, all considerations apply with trivial modifications to the interference terms and the SM 
Drell-Yan background as well. Hence, as our final result, the cross section in \eqn{eq:SigmaHatSplit} 
can be calculated from
\begin{equation}\label{From_Partons_To_Hadrons}
\frac{\d{ \sigma_{ij}}}{\d{\mll}}  =  
\frac{\sqrt{\ps}}{16\pi N_c} \frac{\Bigl(\ps - m_i^2 \Bigr) \rund{\ps - m_j^2 }  +  m_i m_j \Gamma_i
\Gamma_j}{\eck{\Bigl( \ps - m_i^2\Bigr)^2 + m_i^2\Gamma_i^2}  \eck{\rund{\ps - m_j^2}^2 + m_j^2\Gamma_j^2}  }
   	 \sum_q  c^q_{2,ij}  \Tf{}{q,2	} \; 
\end{equation}
with $i,j=\gamma,Z,\Zp$, where
$c^q_{2,ij} =  2 \eck{ g^V_{q,i} g^V_{q,j} + g^A_{q,i} g^A_{q,j} } \cdot 
\eck{ g^V_{l,i} g^V_{l,j} + g^A_{l,i} g^A_{l,j} }$ is given in terms of the couplings of the vector bosons 
to fermions defined for the photon and the $Z$ boson in analogy to \eqn{eq:Lagrangian}. 

\begin{figure}[t]
	\centering    
        \includegraphics[width=0.49\textwidth]{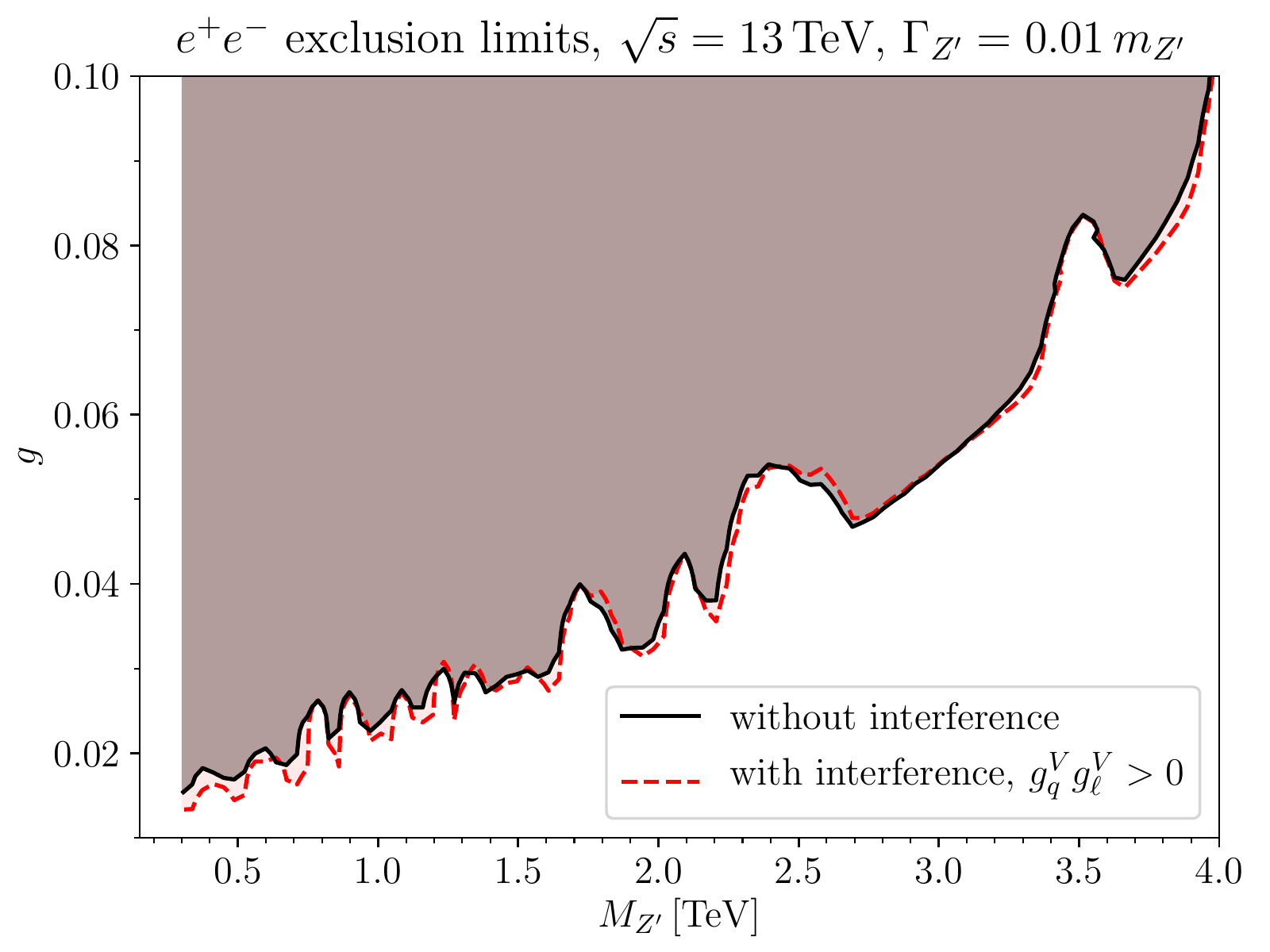}\hfill\includegraphics[width=0.49\textwidth]{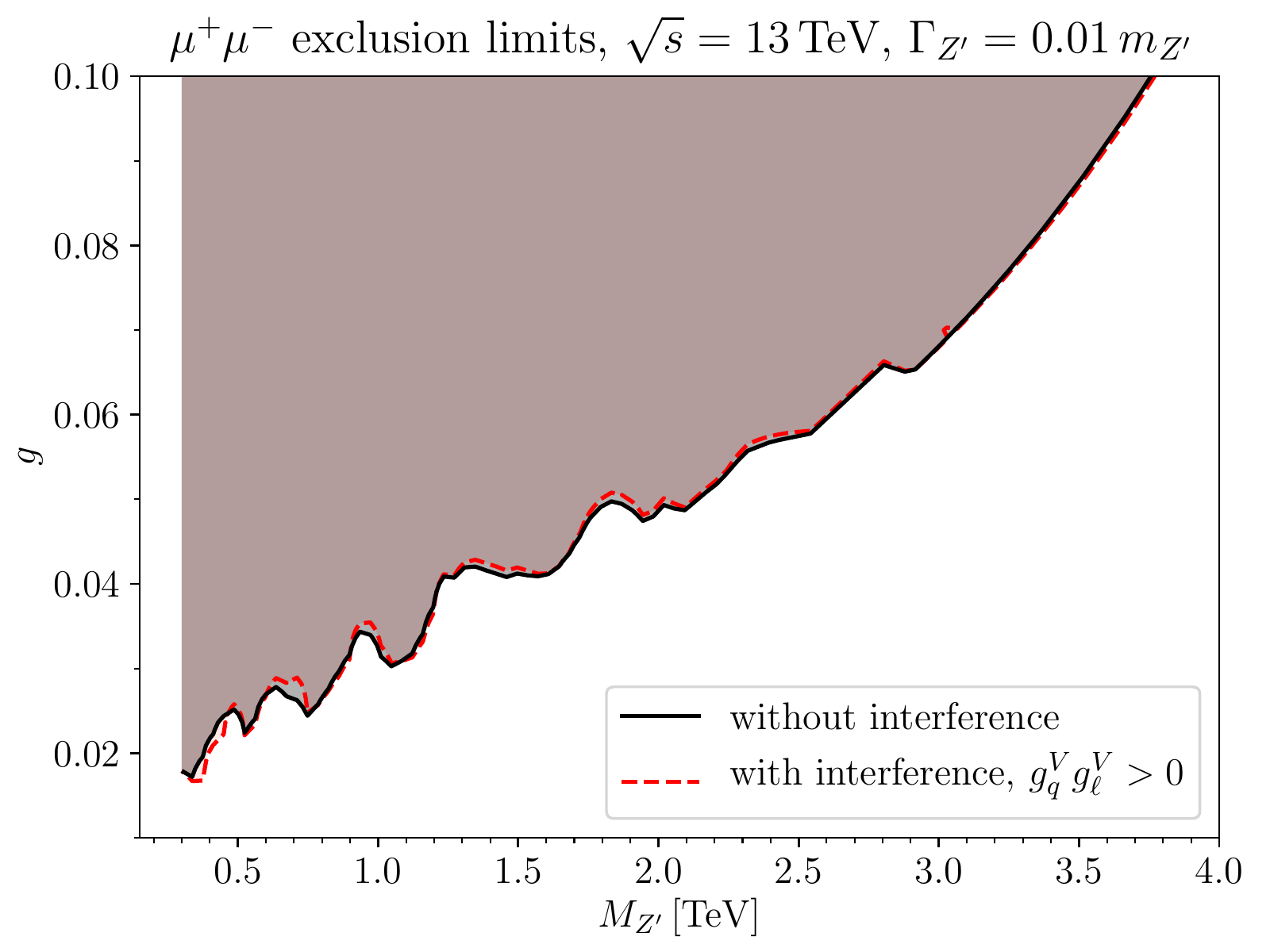}
        \includegraphics[width=0.49\textwidth]{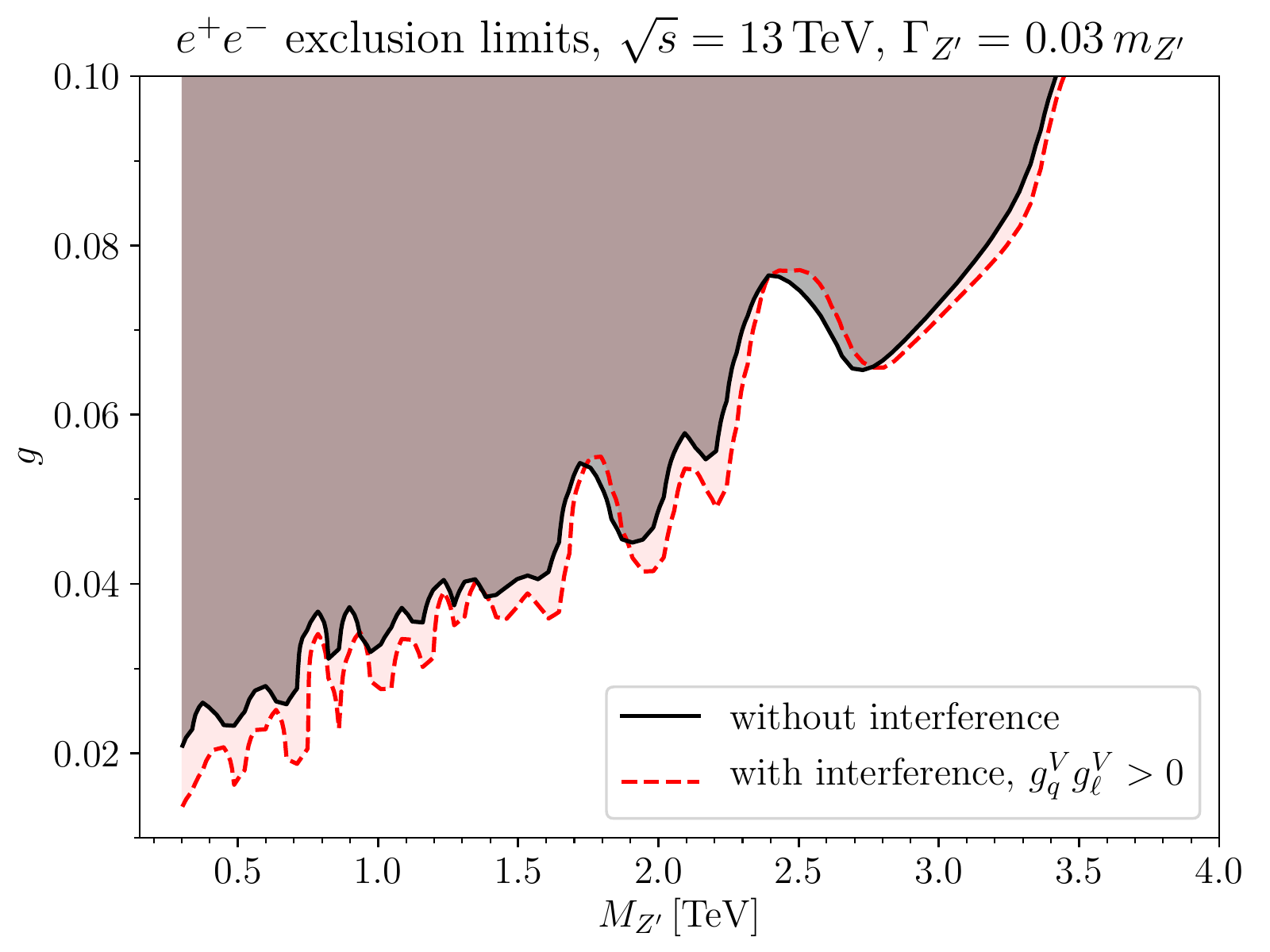}\hfill\includegraphics[width=0.49\textwidth]{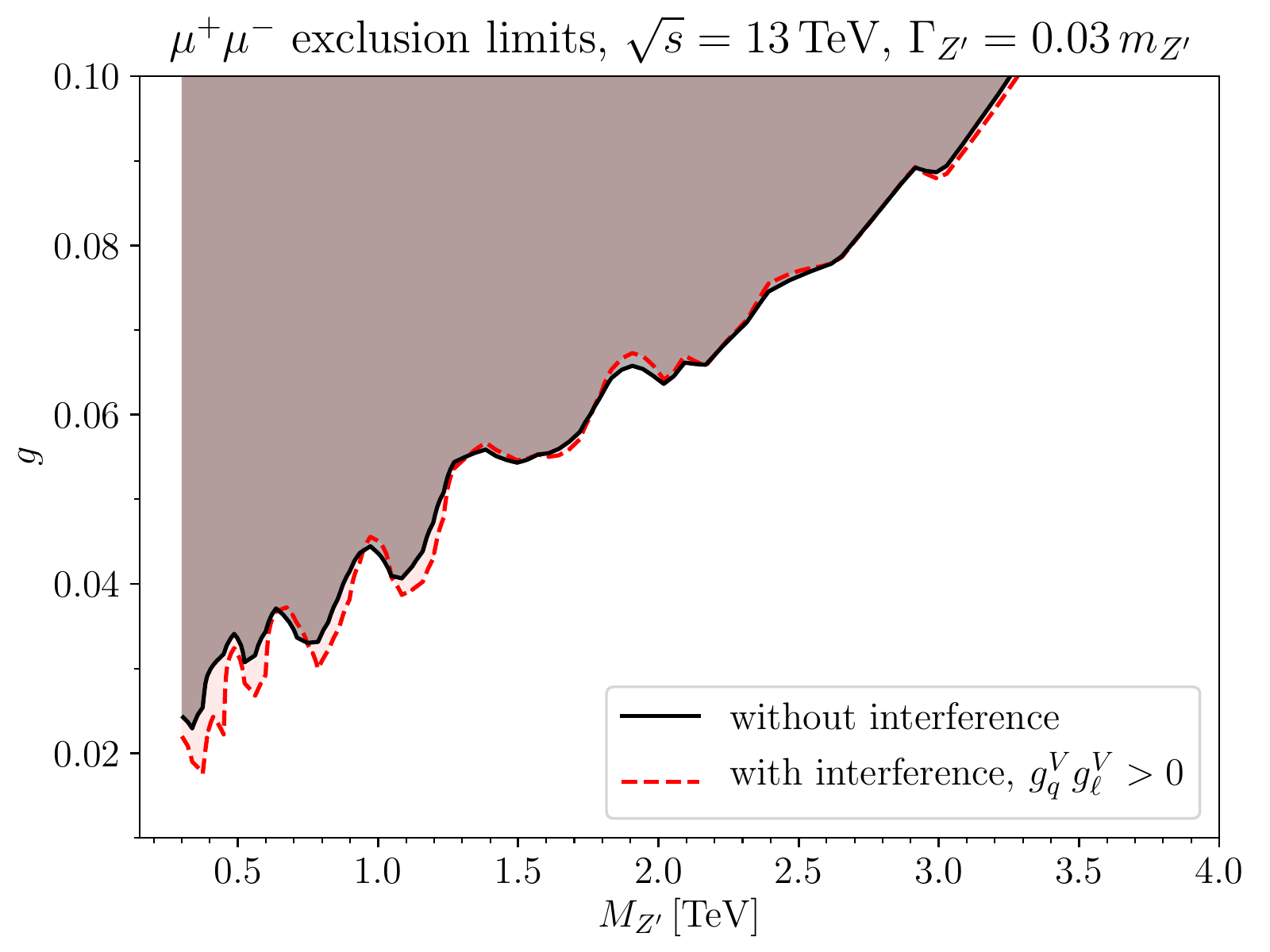}
        \includegraphics[width=0.49\textwidth]{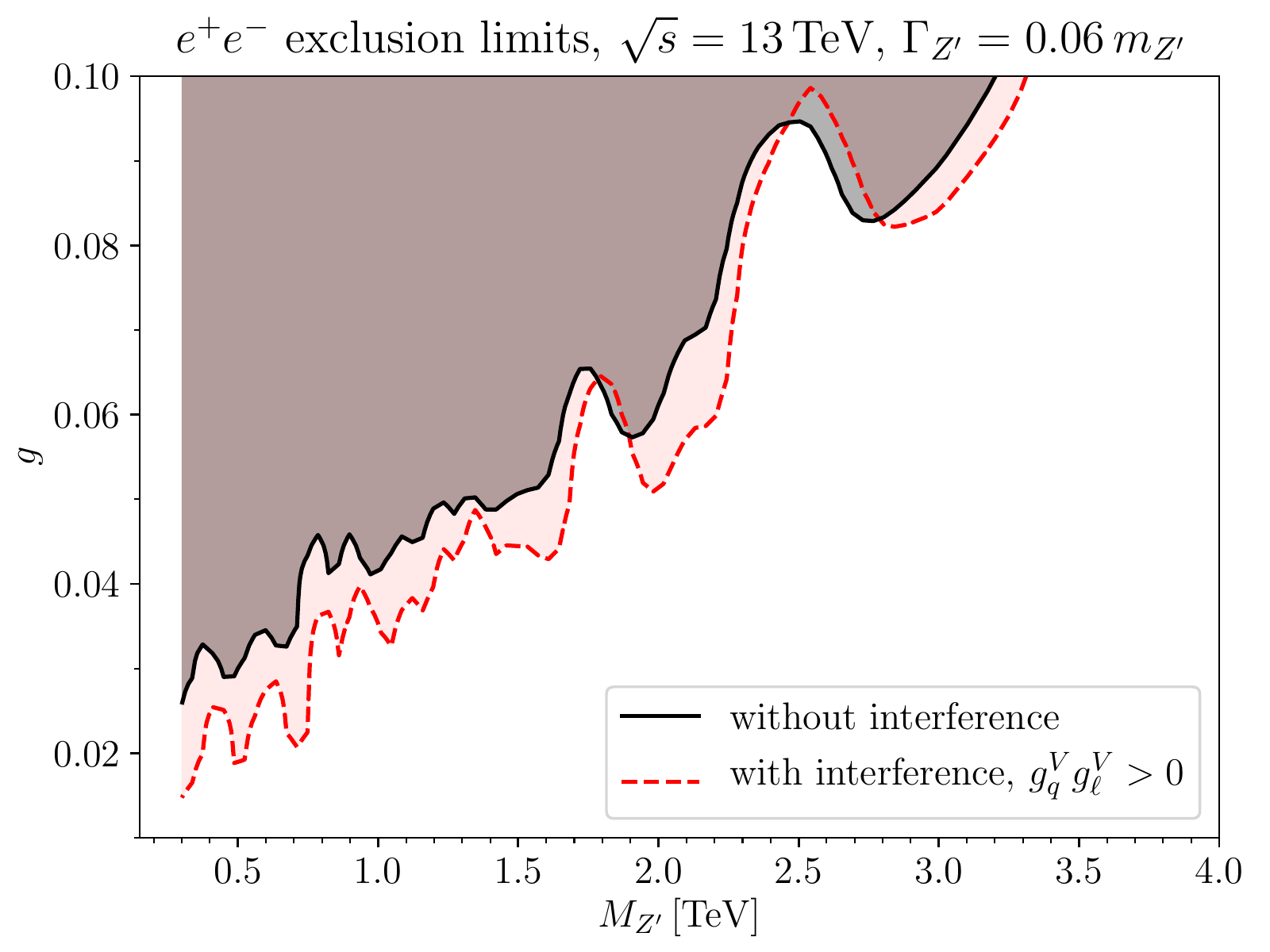}\hfill\includegraphics[width=0.49\textwidth]{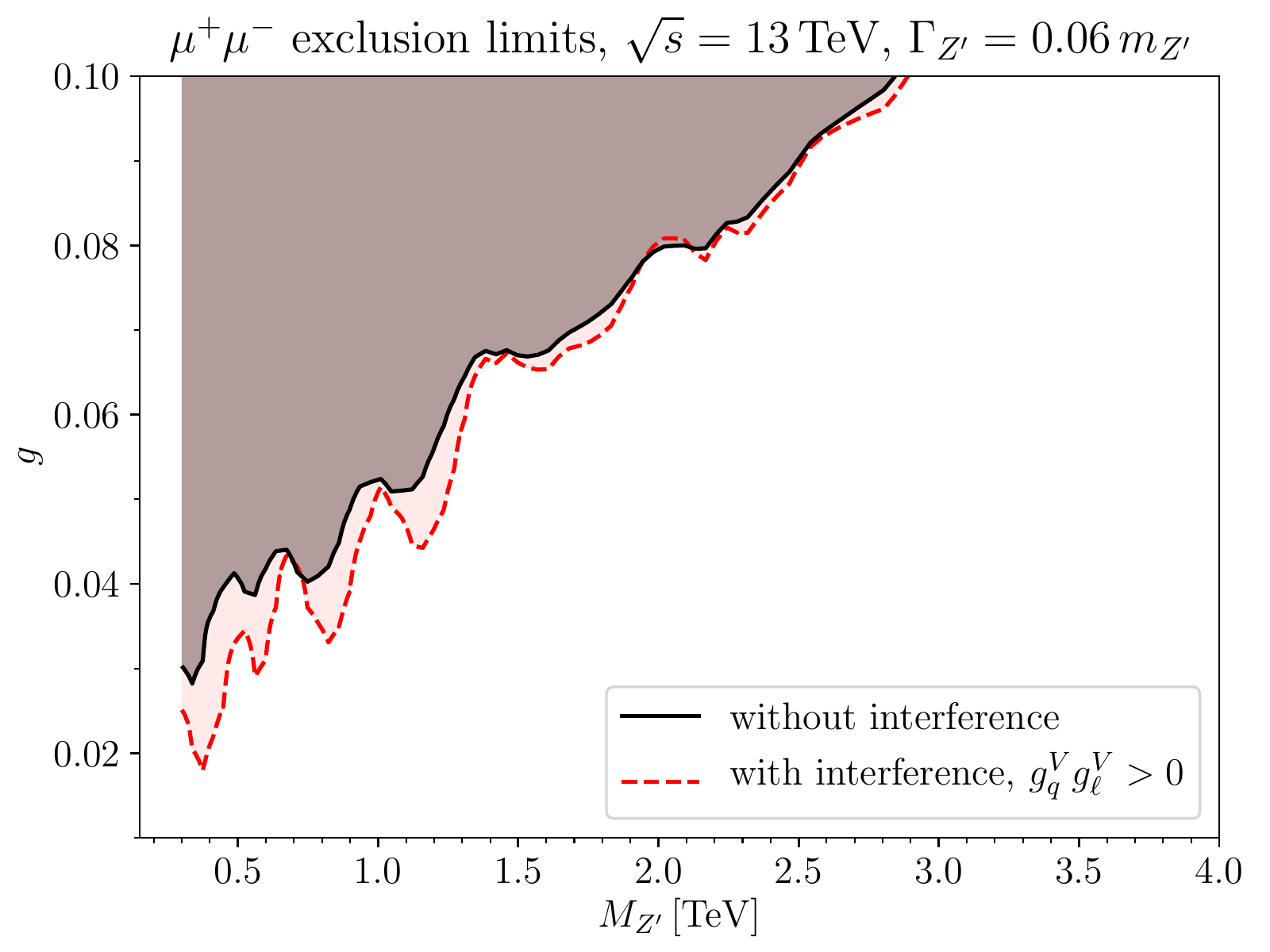}
    \caption{Same as figure~\ref{fig:chi2_couplings} but separately for the effective coupling to electrons (left column) and muons (right column). 
    }\label{fig:chi2_couplings_eemm}
\end{figure}

\begin{figure}[t]
	\centering    
        \includegraphics[width=0.605\textwidth]{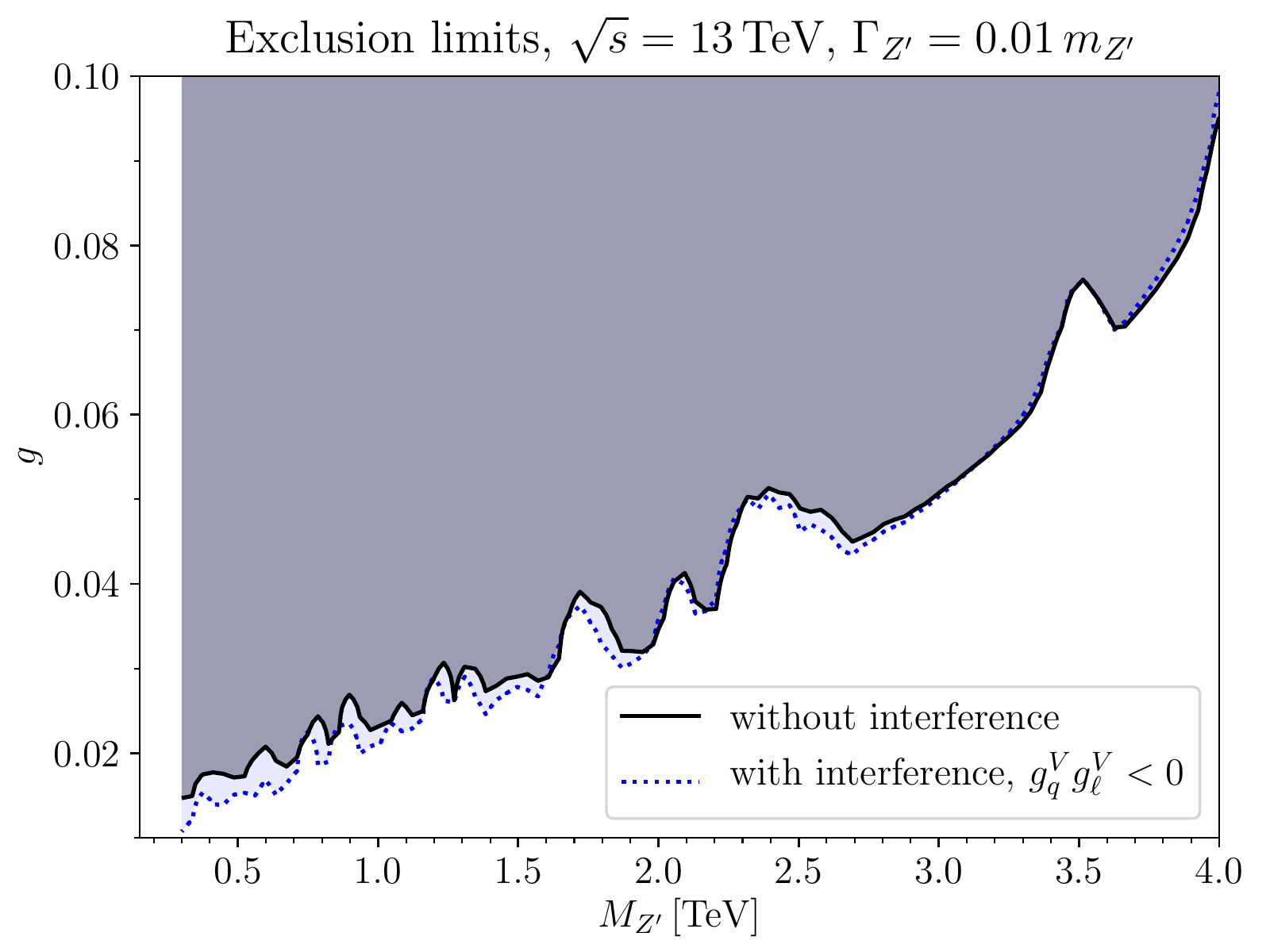} 
        \includegraphics[width=0.605\textwidth]{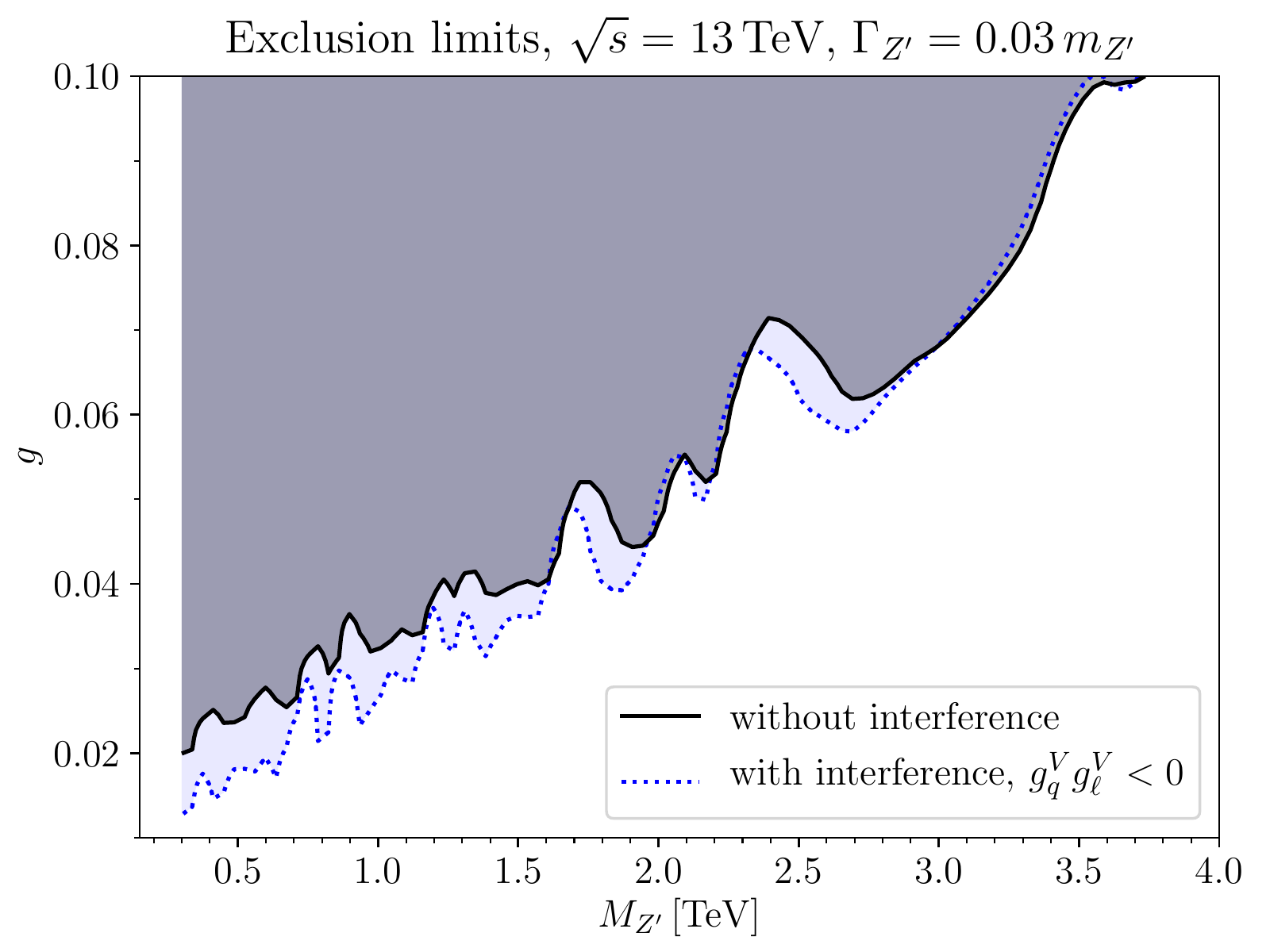}
        \includegraphics[width=0.605\textwidth]{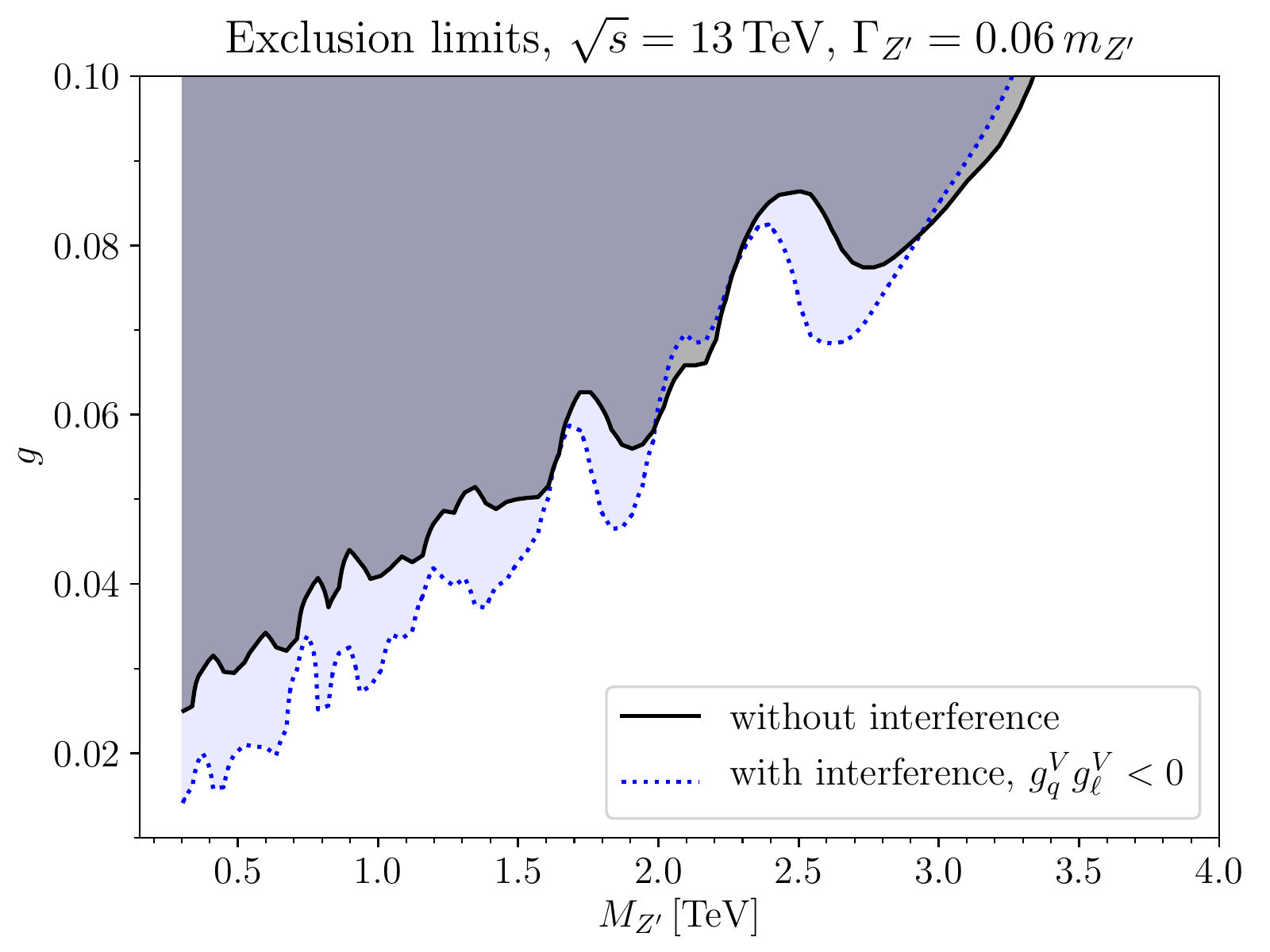}
    \caption{Same as figure~\ref{fig:chi2_couplings} but for the case that $\gqv \glv < 0$. 
    }\label{fig:chi2_couplings_opposite}
\end{figure}

\section{Further Exclusion Limits}
\label{app:opposite}

In figure~\ref{fig:chi2_couplings_eemm} we show separate constraints on the effective coupling $g$ to electrons and muons. We find constraints on the former to be slightly stronger than on the latter, which is a direct consequence of the better detector resolution for the electron final state. Although fluctuations are more pronounced in the electron channel, the effect of interference is similar in both cases.

\begin{figure}[t]
    \centering
    \includegraphics[width= 0.59 \textwidth]{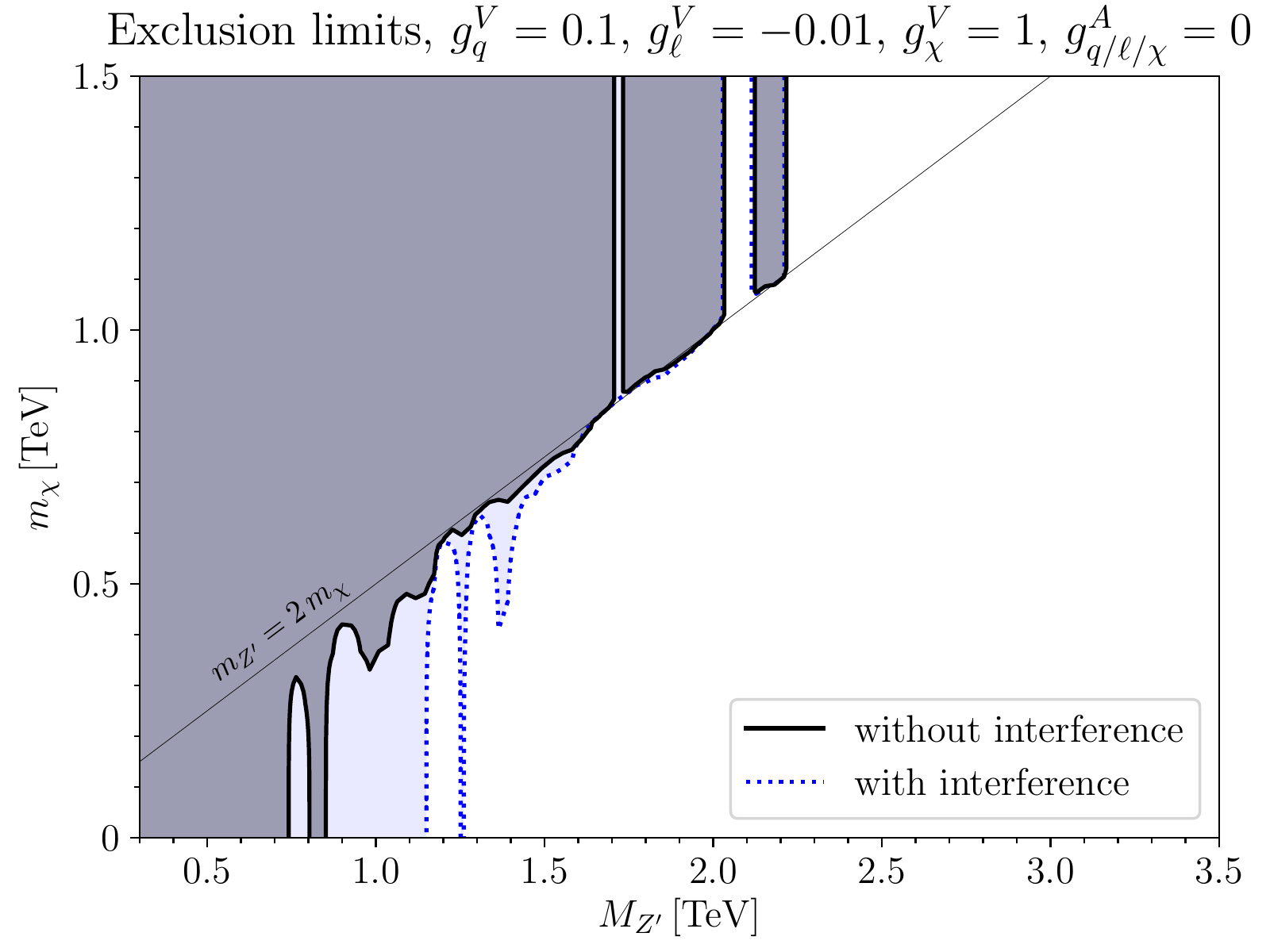} 
    \includegraphics[width= 0.59 \textwidth]{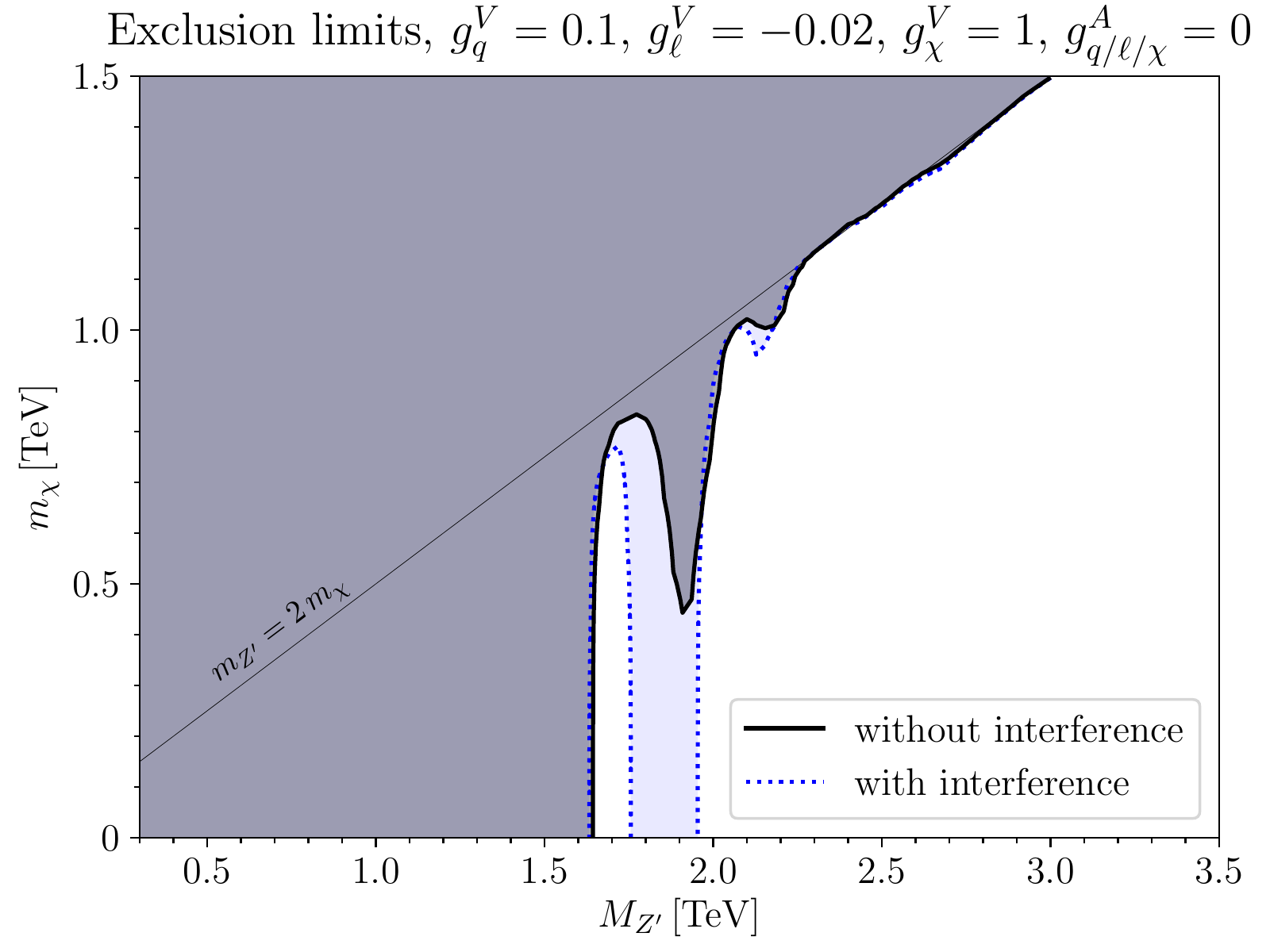}
    \caption{Same as figure~\ref{fig:chi2_mass} but for the case that $\gqv \glv < 0$.
    }\label{fig:chi2_mass_opposite}
\end{figure}

\begin{figure}[t]
	\centering    
        \includegraphics[width=0.49\textwidth]{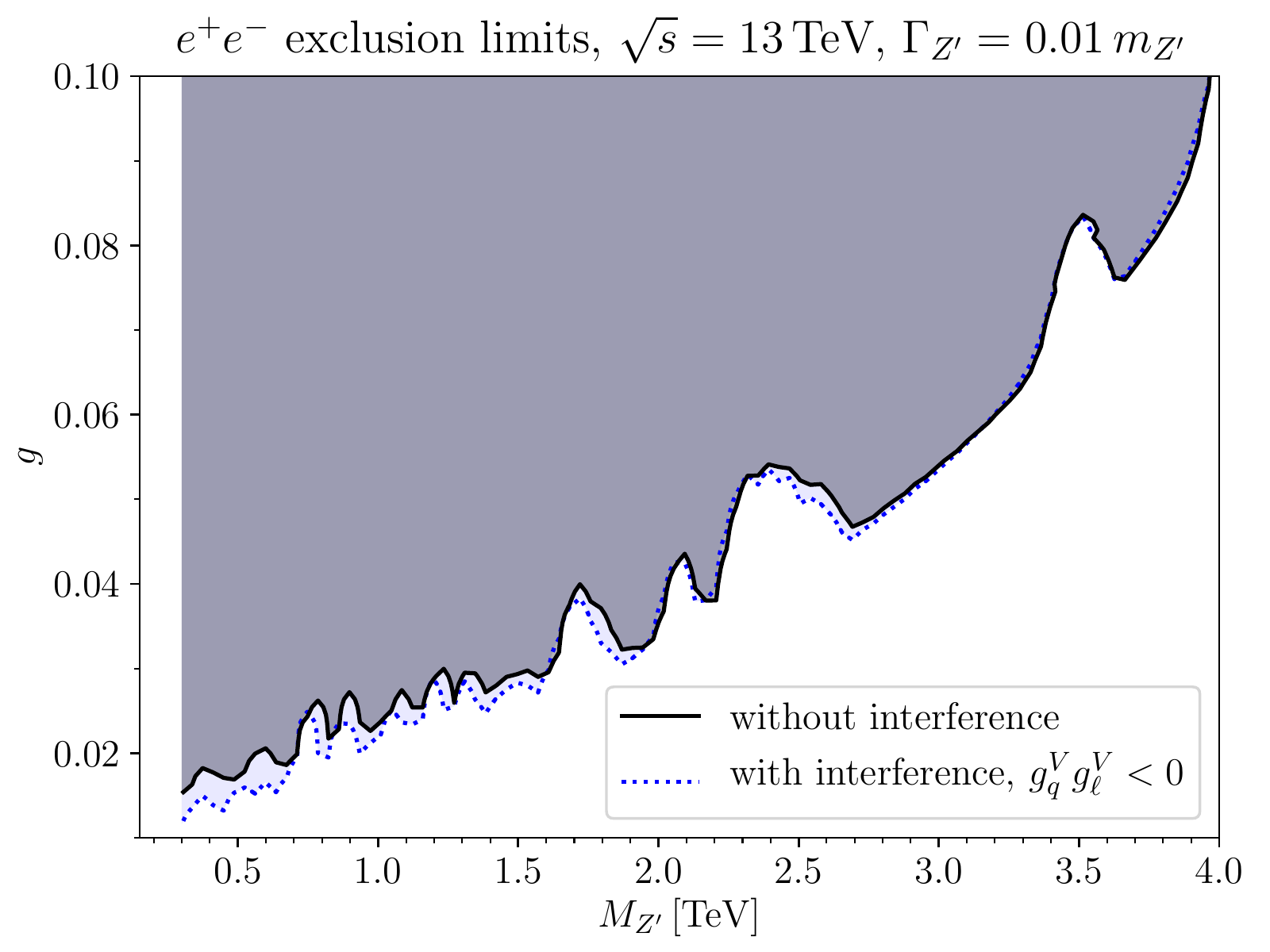}\hfill\includegraphics[width=0.49\textwidth]{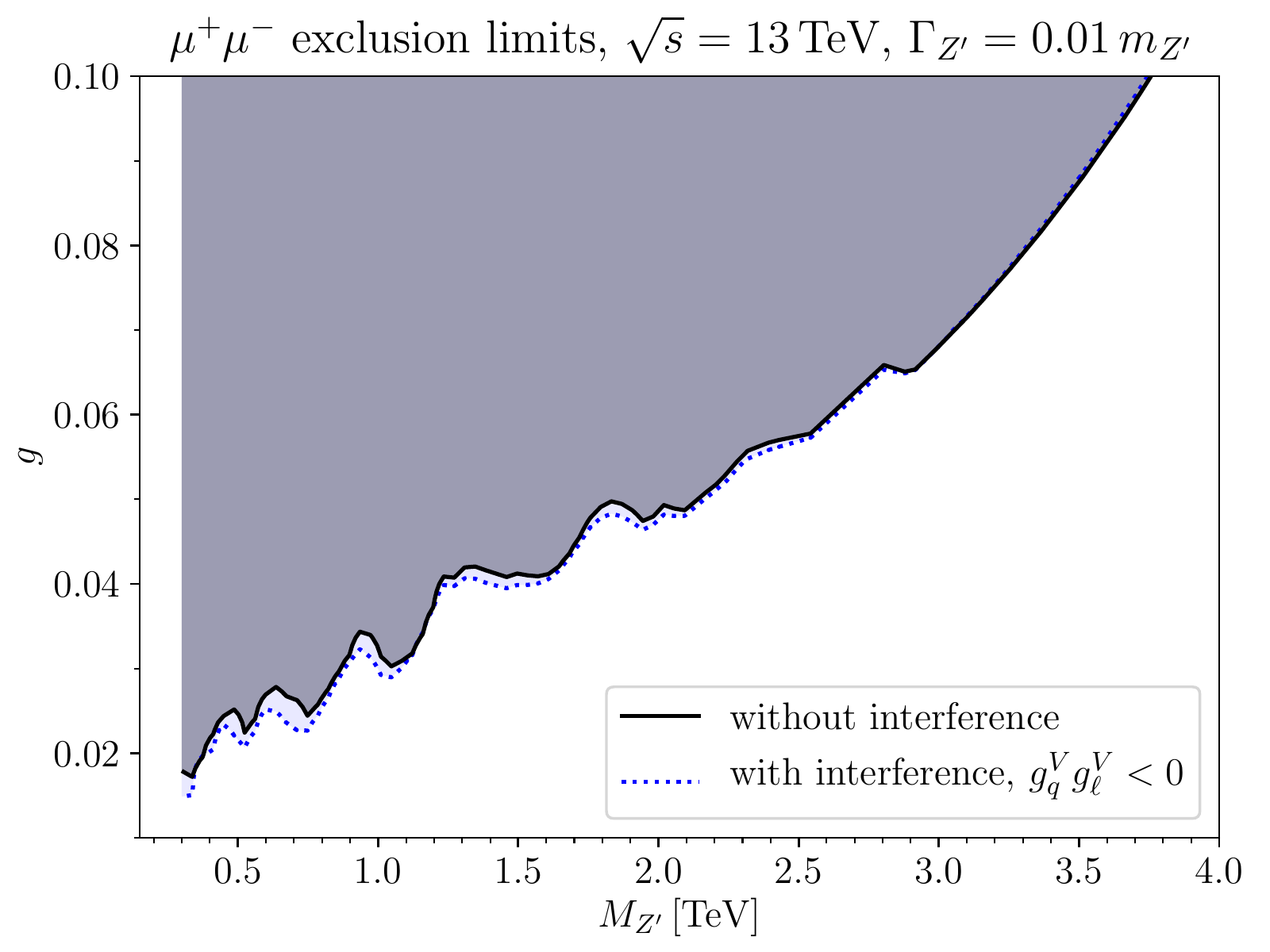}
        \includegraphics[width=0.49\textwidth]{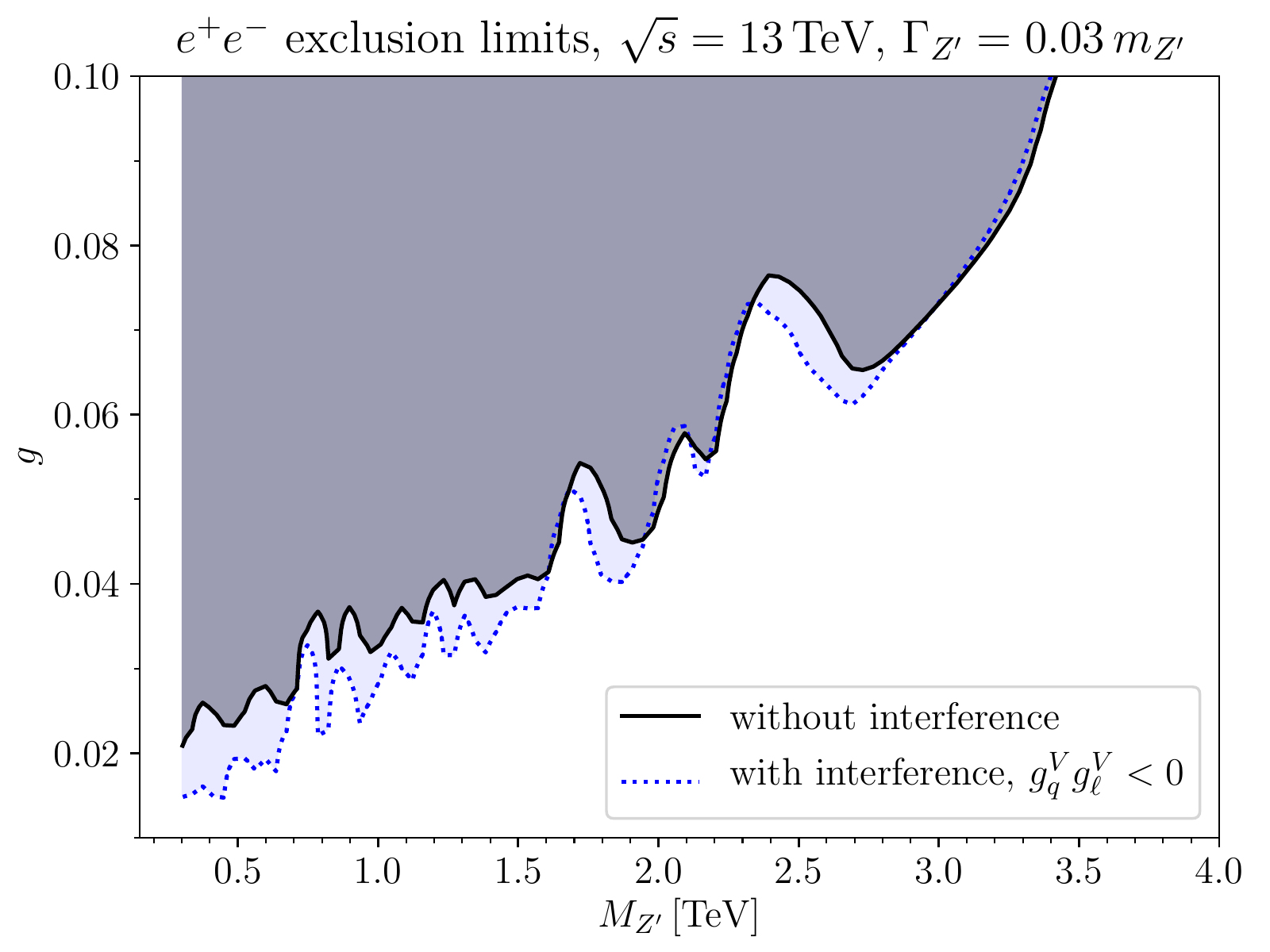}\hfill\includegraphics[width=0.49\textwidth]{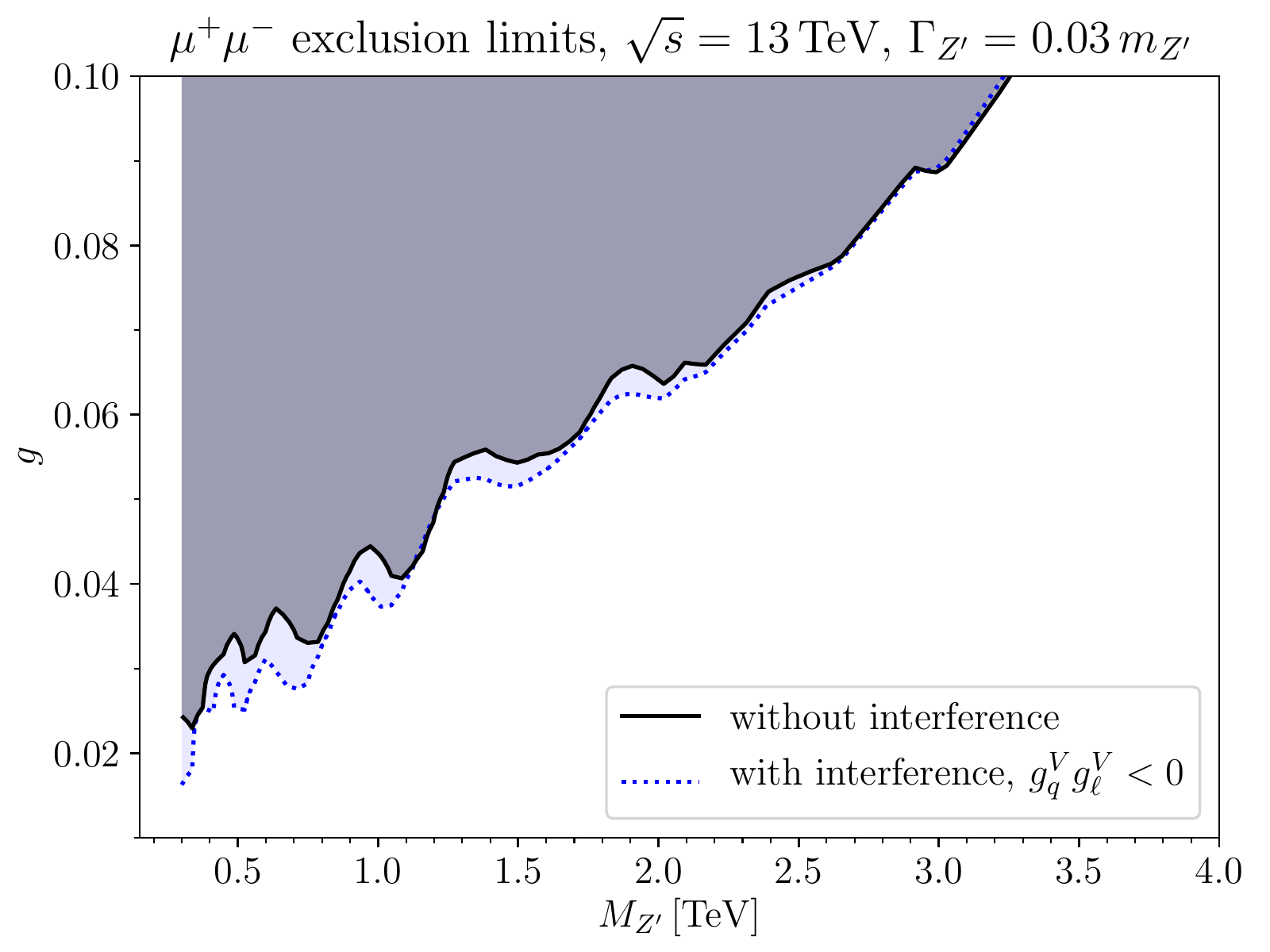}
        \includegraphics[width=0.49\textwidth]{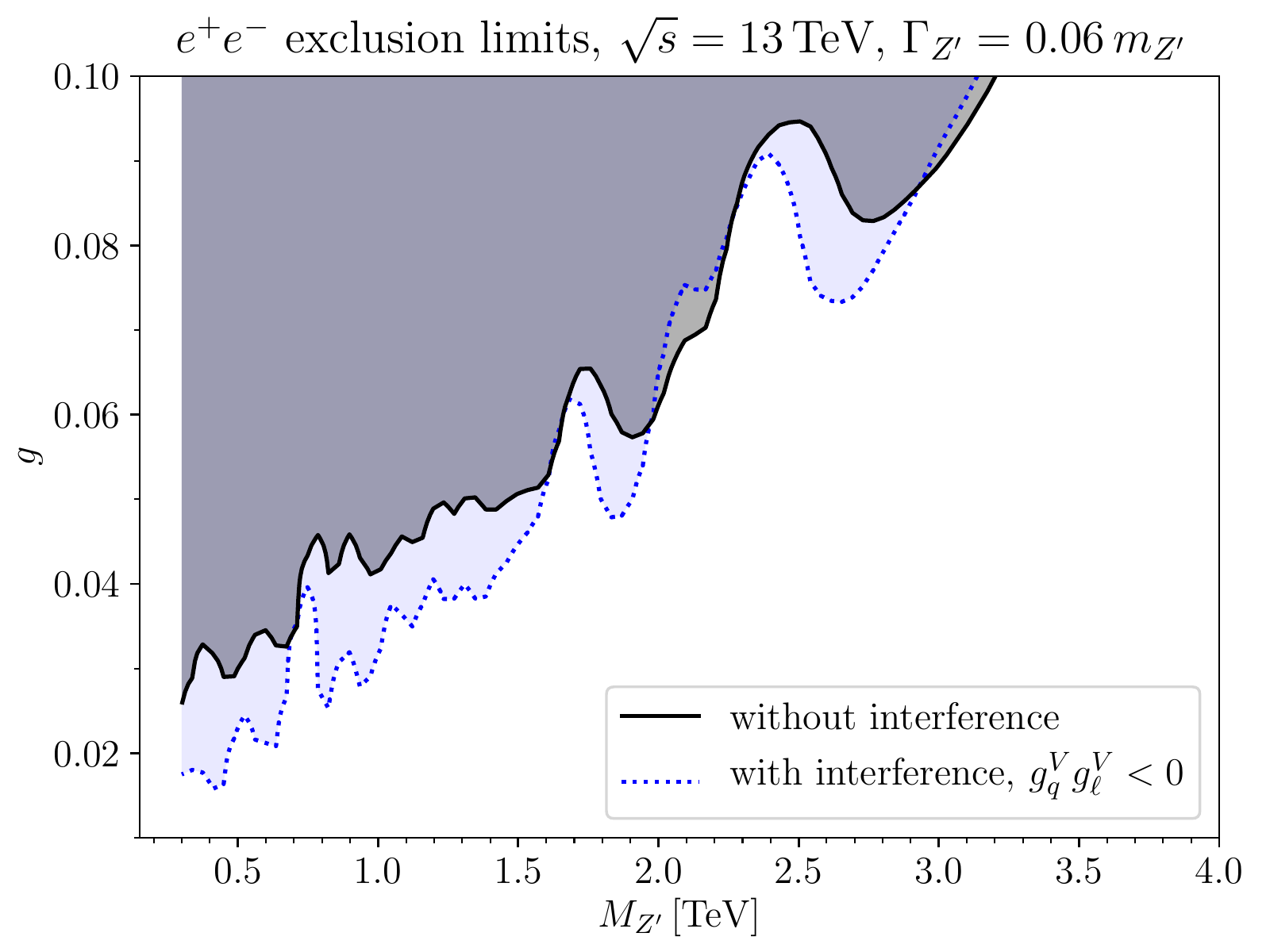}\hfill\includegraphics[width=0.49\textwidth]{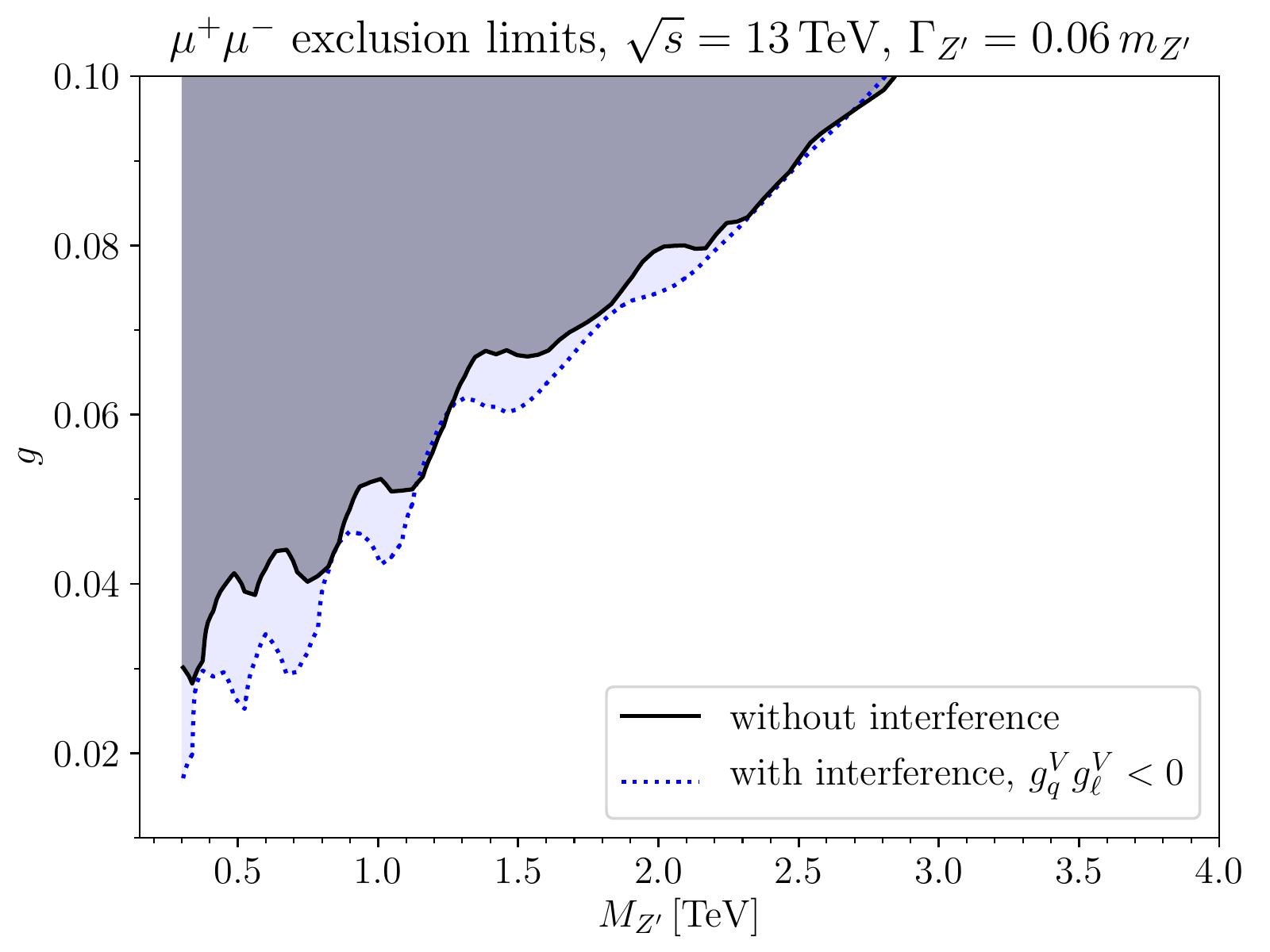}
    \caption{Same as figure~\ref{fig:chi2_couplings_eemm} but for the case that $\gqv \glv < 0$. 
    }\label{fig:chi2_couplings_eemm_opposite}
\end{figure}

In figures~\ref{fig:chi2_couplings_opposite},~\ref{fig:chi2_mass_opposite} and~\ref{fig:chi2_couplings_eemm_opposite} we present our results for the case that the product of quark and lepton coupling are negative ($\gqv \glv < 0$). As can be seen from figure~\ref{fig:InterferenceSignals}, this changes the shape of the expected signal substantially. Crucially, interference still leads to an increased height of the peak and therefore including interference effects typically leads to stronger exclusion bounds.

\providecommand{\href}[2]{#2}\begingroup\raggedright\endgroup

\end{document}